\documentstyle[12pt,epsf]{article}
\newcommand {\beq}{\begin{equation}}
\newcommand {\eeq}{\end{equation}}
\newcommand {\beqa}{\begin{eqnarray}}
\newcommand {\eeqa}{\end{eqnarray}}
\newcommand {\beqan}{\begin{eqnarray*}}
\newcommand {\eeqan}{\end{eqnarray*}}
\newcommand {\n}{\nonumber \\}

\newcommand {\Romannumeral}[1]{\uppercase\expandafter{\romannumeral#1}}

\newcommand {\ee}{\mbox{e}}

\newcommand {\dd}{\mbox{d}}

\newcommand {\del}{\partial}

\def\rref#1{(\ref{#1})}

\begin{document}
\setlength{\oddsidemargin}{0cm}
\setlength{\baselineskip}{7mm}

\begin{titlepage}
 \renewcommand{\thefootnote}{\fnsymbol{footnote}}
    \begin{normalsize}
     \begin{flushright}
                     TIT-HEP-275\\
                     KEK-TH-423\\
                    UT-Komaba/94-22\\
                    December 1994
{}~~
     \end{flushright}
    \end{normalsize}
    \begin{Large}
       \vspace{1cm}
       \begin{center}
         {\LARGE Two--loop Renormalization in Quantum Gravity
                 near Two Dimensions} \\
       \end{center}
    \end{Large}

  \vspace{10mm}

\begin{center}
           Toshiaki A{\sc ida}$^{1)}$\footnote
           {E-mail address : aida@phys.titech.ac.jp},
           Yoshihisa K{\sc itazawa}$^{1)}$\footnote
           {E-mail address : kitazawa@phys.titech.ac.jp},
           Jun N{\sc ishimura}$^{2)}$\footnote
           {E-mail address : nisimura@theory.kek.jp,
{}~JSPS Research Fellow.}{\sc and}
           Asato T{\sc suchiya}$^{3)}$\footnote
           {E-mail address : tsuchiya@hep1.c.u-tokyo.ac.jp}\\
      \vspace{1cm}
        $^{1)}$ {\it Department of Physics,
Tokyo Institute of Technology,} \\
                 {\it Oh-okayama, Meguro-ku, Tokyo 152, Japan}\\
        $^{2)}$ {\it National Laboratory for High Energy Physics (KEK),}\\
               {\it Tsukuba, Ibaraki 305, Japan}\\
        $^{3)}$ {\it Institute of Physics, University of Tokyo,} \\
              {\it Komaba, Meguro-ku, Tokyo 153, Japan}\\
\vspace{15mm}

\end{center}
\begin{abstract}
\noindent
We study two--loop renormalization in $(2+\epsilon)$--dimensional
quantum gravity.
As a first step towards the full calculation,
we concentrate on the divergences which are proportional to the
number of matter fields.
We calculate the $\beta$ functions and
show how the nonlocal divergences as well as the infrared divergences
cancel among the diagrams.
Although the formalism includes a subtlety concerning the general covariance
due to the dynamics of the conformal mode,
we find that the renormalization group allows the existence of a fixed point
which possesses the general covariance.
Our results strongly suggest that we can construct a consistent theory
of quantum gravity by the $\epsilon$ expansion around two dimensions.
\end{abstract}
\end{titlepage}
\vfil\eject

\setcounter{footnote}{0}

\section{Introduction}
\setcounter{equation}{0}
Quantum gravity beyond two dimensions may be renormalizable in the
$2+\epsilon$
expansion approach. The remarkable point is that we find the short distance
fixed point in the renormalization group for proper matter contents.
Therefore the gravitational interaction may not become uncontrollably strong
at short distance in quantum theory \cite{2+epsilon,KN,KKN1,KKN2,AKKN}.
Let us consider the matter scattering due to gravitation. Since the
gravitational
coupling constant (Newton's constant) has a dimension, the cross section grows
at short distance and ultimately exceeds the unitarity bound.
On the other hand,
we can define the dimensionless gravitational coupling constant $G$
by introducing the renormalization scale $\mu$ in quantum theory.
If the dimensionless gravitational
coupling constant $G$ possesses a short distance fixed point, the unitarity
problem can be overcome since the cross section at the momentum scale
$p^2 = \mu ^2$ is $f(G(\mu))/{\mu ^2}$
where $f$ is a calculable function of $G$,
on dimensional grounds.

Needless to say, unitarity might be broken by other sources such as
black holes in quantum gravity and there might be a real paradox here.
We hope to address these questions also in the $2+\epsilon$ expansion of
quantum
gravity.
Although a consistent quantum theory of gravitation may require more than local
field theory such as superstring in four dimensions, we should not forget
this simpler possibility.
At least we can learn lessons of quantum gravity
in this simple setting in low dimensions.
This approach is also useful to study two--dimensional quantum gravity
and string theory \cite{NTT1,NTT2,KST}.

As is widely perceived, the renormalization group is one of the most
powerful tool to study quantum field theory. In quantum gravity, we need
to examine the meaning of the renormalization group carefully
since the
spacetime distance itself fluctuates.
Let us consider the Einstein action in $D=2+\epsilon$ dimensions
\beq
\int \dd^{D} x{1\over G_0} \sqrt{g}  R ,
\label{Ein1}
\eeq
where $G_0$ is the gravitational coupling constant.

We parametrize the metric $g_{\mu\nu}$ as $g_{\mu\nu}=\tilde{g} _{\mu\nu}
\ee^{-\phi}$ where $\phi$ is the conformal mode of the metric.
$\tilde{g}_{\mu\nu}$ can be parametrized by a traceless symmetric tensor
$h_{\mu\nu}$ as $\tilde{g}_{\mu\nu} = (\ee^h)_{\mu\nu}$.
We further introduce the dimensionless
gravitational coupling constant $G$ by introducing the renormalization
scale $\mu$ by ${1\over {G_0}} = {{{\mu}^{\epsilon}}\over G}$.
In this parametrization, our action becomes
\beq
\int \dd^{D} x {{{\mu}^{\epsilon}} \over G}
\ee^{-{\epsilon \over 2} \phi} \left( \tilde{R} -
{{\epsilon (D-1)}\over 4} \tilde{g}^{\mu\nu}
\partial _{\mu} \phi \partial _{\nu} \phi \right)  ,
\label{Ein2}
\eeq
where $\tilde{R}$ is the scalar curvature made out of $\tilde{g}_{\mu\nu}$.
We note the similarity of this action to the nonlinear sigma model
\beq
\int \dd^D x {{{\mu}^{\epsilon}} \over F}
\ee^{-{\epsilon \over 2} \phi}
\partial _{\mu} \vec{n} \cdot \partial ^{\mu} \vec{n} .
\eeq
Since $\tilde{R}$ involves two derivatives, it is analogous to
the kinetic term of the nonlinear sigma model. $F$ is the coupling constant
for the $\vec{n}$ field and $G$ is that for the $h_{\mu\nu}$ field.
The physical length scale is set by the line
element $ds^2=\ee^{-\phi}dx_\mu dx^\mu$.
If we scale the length as $ds^2 \rightarrow \lambda^2 ds^2$,
the coupling constant changes as ${1\over F} \rightarrow
{{\lambda ^{\epsilon}}\over F}$ and ${1\over G} \rightarrow
{{\lambda ^{\epsilon}}\over G}$ respectively.
In this sense the coupling constants grow canonically at
short distance in both theories.
However we can choose the renormalization scale $\mu$ such
that $\mu \lambda = 1$ and consider the running coupling
constants.
If the running coupling constants possess the short distance
fixed points, the theory is under control.

The novel feature of quantum gravity is that the zero mode of
the $\phi$ field
sets the scale of the metric
and hence the scale of the length.
In fact the zero mode is determined by the classical solution
of the theory.
For example
the scale of the metric expands with time in our universe.
Therefore we can take the constant mode of $\phi$ to be the present scale
factor of the metric.
Since the definite combination ${\mu}^{\epsilon} \ee^{-{\epsilon\over 2}\phi}$
appears in the action, it is most advantageous to choose the renormalization
scale $\mu$ to compensate the scale factor of the metric (or the constant
mode of $\phi$). It is analogous to choose the renormalization scale
to match the momentum scale of the relevant scattering in the conventional
field theory problem. In this way the renormalization scale
of the dimensionless
gravitational coupling constant $G(\mu )$ is related to the scale factor
of the metric. In particular, large renormalization scale is
relevant at short distance physics.

We need to consider all possible values of the constant mode of $\phi$
for the whole theory
since we are integrating over it. Therefore we consider the whole
renormalization group trajectory as the whole quantum theory of
gravitation. Such an idea satisfies the
independence of
the theory from the scale factor of a particular metric.
In our universe, the scale factor of the metric can be identified
with time. In this interpretation of the renormalization group
in quantum gravity, we may say that
the renormalization scale is identified with time.
The renormalization group evolution is hence naturally related to the
time evolution in quantum gravity.

In this paper we study the two--loop renormalization of Einstein gravity
coupled to $c$ copies of scalar fields in the conformally invariant way with
the action
\beq
{\mu ^\epsilon\over G}\int \dd^D x \sqrt{g}
\left\{R \left( 1-{\epsilon \over {8(D-1)}}\varphi ^2_i \right)
+{1\over 2}g^{\mu\nu}\partial _{\mu} \varphi _i\partial _{\nu} \varphi _i
\right\}
\label{action1} ,
\eeq
where $i$ runs from $1$ to $c$.
This action can be rewritten as
\beq
{\mu ^\epsilon\over G}\int \dd^D x \sqrt{\hat{g}}
\left[ \tilde{R} \left\{  \left( 1+{1\over 2}
\sqrt{\epsilon \over {2(D-1)}}\psi \right) ^2
-{\epsilon \over {8(D-1)}}\varphi ^2_i \right\}
-{1\over 2}\tilde{g}^{\mu\nu}\partial _{\mu} \psi\partial _{\nu} \psi
+{1\over 2}\tilde{g}^{\mu\nu}\partial _{\mu} \varphi _i\partial _{\nu}
\varphi _i  \right] ,
\label{action2}
\eeq
where we have reparametrized the conformal mode as
$\ee^{-{\epsilon \phi\over 4}} =
\lambda ^{\epsilon \over 2}
\left( 1+{1\over 2} \sqrt{\epsilon \over {2(D-1)}} \psi \right)$
in order to make the kinetic
term of $\psi$ canonical. The $\lambda$ factor can be cancelled by choosing
the appropriate renormalization scale $\mu$.
In this way we can get rid of the $1\over \epsilon$
pole of the conformal mode propagator. Note that the conformal mode $\psi$
can be viewed as another conformally coupled scalar field
in this parametrization.
Therefore we can quantize the theory treating the conformal mode as a matter
field coupled in the conformally invariant way. In such a quantization
procedure it is important to keep the conformal invariance. Since it is well
known that the conformal anomaly arises in quantum field theory, we
need to modify the tree action to cancel the quantum conformal anomaly.

It has been proposed to generalize the action in the following form which
possesses the manifest volume--preserving
diffeomorphism invariance \cite{KKN2,AKKN}
\beq
{\mu ^\epsilon\over G}\int \dd^D x \sqrt{\hat{g}}
\left\{ \tilde{R}L(\psi ,\varphi _i )
-{1\over 2}\tilde{g}^{\mu\nu}\partial _{\mu} \psi\partial _{\nu} \psi
+{1\over 2}\tilde{g}^{\mu\nu}\partial _{\mu} \varphi _i\partial _{\nu}
\varphi _i \right\} ,
\label{action3}
\eeq
where $L=1+a\psi + b(\psi ^2 - \varphi _i^2)$.
It has been shown that the theory is renormalizable to the one--loop level
and the $\beta$ functions of the couplings are found to be
\beq
\beta _{G} =\epsilon G - AG^2 ,\;
\beta _a = -{AG\over 2}a ,\;
\beta _b = 0,
\label{beta1}
\eeq
where $A={25-c\over {24\pi}}$.
The Einstein action is the infrared fixed point with $G=0,a=\sqrt{\epsilon\over
{2(D-1)}}$ and $b={\epsilon \over {8(D-1)}}$. The theory possesses the short
distance fixed point with $G={\epsilon \over A}, a=0$ and
$ b= {\epsilon \over {8(D-1)}}$.
The conformal anomaly is shown to be cancelled out on the whole
renormalization group trajectory.

It is important to perform the two--loop renormalization of the
theory. It serves to establish the validity of the $2+\epsilon$ expansion
in quantum gravity by showing that the higher order corrections can
be computed systematically. However the two--loop calculations in quantum
gravity is a formidable task due to the  proliferation of diagrams and
tensor indices. Therefore we have decided to calculate the two--loop
counterterms which are proportional to the number of matter fields
(the central charge) first. In this paper we report the result of
such a calculation. Since the number of scalar fields we couple to
gravity is a free parameter, the counterterms must be of the
renormalizable form. They further must satisfy the requirement
from the general covariance.
Therefore this calculation serves as a check of
the $2+\epsilon$ expansion approach.

This paper is organized as follows. In section 2, we briefly review the
one--loop renormalization of the action \rref{action3} .
In section 3, we explain our two--loop calculation of the counterterms.
In section 4, we state the results of our calculation.
In section 5, we compute the $\beta$ functions and
check the general covariance at the ultraviolet fixed point.
We discuss the physical implications and draw conclusions in section 6.

\vspace{1cm}

\section{Brief Review on the  One--loop Renormalization}
\setcounter{equation}{0}

We utilize the background field method to compute the effective action.
The generating functional for the connected Green's functions
in the field theory is
\beq
\ee^{-W[J]} = \int {\cal D} \varphi ~\exp(-S -J\cdot \varphi),
\label{W}
\eeq
where $S$ is the action and  $J\cdot \varphi \equiv
\int \dd^D x ~J(x)\varphi(x)$. $\varphi$ denotes a collection of
fields in the theory.
In this paper the metric is taken to be Euclidean since the Euclidean
rotation from the Minkowski metric is straightforward within the perturbation
theory.

The effective action is obtained by the Legendre transform
\beq
\Gamma [\langle\varphi \rangle]= W[J] - J\cdot \langle\varphi \rangle,
\label{Gamma}
\eeq
where $\langle\varphi (x)\rangle={{\delta W[J]}\over {\delta J (x)}}$.
Therefore the effective action is
\beq
\ee^{-\Gamma [\langle\varphi\rangle]} = \int {\cal D} \tilde{\varphi}
{}~\exp \left( - S[\langle\varphi\rangle+ \tilde{\varphi}]
+{{\delta \Gamma [\langle\varphi\rangle]}\over
{\delta \langle\varphi\rangle}}\cdot \tilde{\varphi}\right) ,
\label{Gamma2}
\eeq
where $\tilde{\varphi} = \varphi - \langle\varphi\rangle$
since $J=-{{\delta \Gamma}\over
{\delta \langle\varphi \rangle}}$.
The effective action can be expanded in terms of $\hbar$ as
\beq
\Gamma = S + \hbar \Gamma^{(1)} + \hbar ^2 \Gamma^{(2)} + \cdots .
\eeq
Hence we can compute the effective action by expanding the action $S$
around the background $\langle\varphi \rangle$ and dropping the linear terms in
$\tilde{\varphi}$.
Namely the effective action is the sum of the one--particle
irreducible diagrams with respect to $\tilde{\varphi}$.


In our context, we decompose the fields into the backgrounds and the
quantum fields as  $\varphi _i \rightarrow \hat{\varphi}_i + \varphi _i$,
$\psi \rightarrow \hat{\psi} + \psi$ and
$\tilde{g}_{\mu\nu} = \hat{g}_{\mu\rho} (\ee^h)^{\rho}_{~\nu}$,
where $h^{\mu}_{~\nu}$ is a traceless symmetric tensor.
The effective action can be computed by summing the one--particle
irreducible diagrams with respect to the quantum fields
$\varphi _i$, $\psi$ and $h^{\mu}_{~\nu}$.


The crucial local gauge invariance
(general covariance) of the action
\rref{action3} in this parametrization is
\begin{eqnarray}
\delta \tilde{g}_{\mu\nu} & = & \partial _{\mu}\epsilon ^{\rho}
\tilde{g}_{\rho\nu} + \tilde{g}_{\mu\rho}\partial _{\nu}\epsilon^{\rho}
+\epsilon^\rho\partial_\rho\tilde{g}_{\mu\nu}
-{2\over D}\partial _{\rho}\epsilon ^{\rho}\tilde{g}_{\mu\nu}, \nonumber \\
\delta \psi & =  & \epsilon ^{\rho} \partial _{\rho} \psi +
(D-1) {{\partial L}
\over {\partial \psi}}{2\over D} \partial _{\rho} \epsilon ^{\rho},
\nonumber \\
\delta \varphi _i & = & \epsilon ^{\rho} \partial _{\rho} \varphi _i -
(D-1) {{\partial L}
\over {\partial \varphi _i}}{2\over D} \partial _{\rho} \epsilon ^{\rho}.
\label{gauge}
\end{eqnarray}

In addition to expanding the action around the background fields,
we need to fix the gauge invariance \rref{gauge} in order to
perform the functional integration.
We adopt the following
background gauge.
\beq
{1\over 2} L
\left(\nabla ^{\mu} h_{\mu\nu} -{{\partial _{\nu} L}\over L}\right)
\left(\nabla _{\rho} h^{\rho\nu} -{{\partial ^{\nu} L}\over L}\right) .
\label{gaugefixing}
\eeq
Throughout this paper the tensor indices are raised and lowered by the
background metric and the covariant derivatives are taken with
respect to the background metric.
We also add the corresponding ghost terms
\beq
\nabla _{\mu}\bar{\eta}_{\nu}\nabla ^{\mu}\eta ^{\nu}
+ \hat{R}_{\nu}^{\mu}\bar{\eta}_{\mu}\eta ^{\nu}-
{{\partial _{\nu}L}\over L}(\nabla ^{\mu}\bar{\eta}_{\mu})
\eta ^{\nu} + \cdots .
\label{ghostaction}
\eeq
The background gauge has the advantage to keep the manifest
general covariance with
respect to the background metric.

The one--loop counterterm of this theory is evaluated to be
\beq
-{A\over \epsilon} \int \dd^D x \sqrt{\hat{g}}\tilde{R} .
\eeq
The resulting $\beta$ functions are quoted in \rref{beta1}.
The conformal anomaly of the theory is found to be
\begin{eqnarray}
& & \left[ \epsilon L -AG-2(D-1)\left\{ \left(
{{\partial L}\over {\partial \psi}} \right)^2
-\left({{\partial L}\over {\partial \varphi _i}}\right)^2\right\}
\right]\tilde{R}
\nonumber \\
&-& {1\over 4} \left\{\epsilon -4(D-1){{\partial ^2 L}\over
{\partial \psi ^2}} \right\} \partial _{\mu} \psi \partial ^{\mu}\psi
\nonumber \\
& +& {1\over 4} \left\{\epsilon +4(D-1){{\partial ^2 L}\over
{\partial \varphi_i ^2}} \right\}\partial _{\mu} \varphi _i
\partial ^{\mu}\varphi ^i .
\label{anomaly}
\end{eqnarray}
It has been shown that the conformal anomaly vanishes
along the renormalization group trajectory.
Note that the conformal invariance is crucial to
restore the general covariance from the action
which possesses only the volume--preserving diffeomorphism
invariance. Therefore the general covariance is also maintained
along the renormalization group trajectory.

We are particularly interested in the short distance fixed
point of the renormalization group. At the fixed point the
tree action is
\beq
{\mu ^{\epsilon}\over G}\int \dd^D x \sqrt{\hat{g}}
\left\{ \tilde{R}L(X_i )
+{1\over 2}\tilde{g}^{\mu\nu}\partial _{\mu} X _i\partial _{\nu}
X ^i \right\} ,
\label{faction}
\eeq
where $L=1- bX _i X^i$. $X_i$ denotes the $\psi$ and $\varphi _i$
fields ($X_0 =\psi$) and $X_i X^i = -\psi ^2 +
\sum_{j=1}^c {\varphi _j}^2$.
This action possesses the $Z_2$ symmetry $X_i \rightarrow
-X_i$.
The enhancement of the symmetry may be due to the fact that we
are expanding the theory around the symmetric vacuum
in which the expectation value of the metric vanishes.
In this paper, we compute the two--loop counterterms
which are proportional to the number of scalar fields $c$.
We compute the counterterms at the fixed point where further
simplification takes place due to the $Z_2$ symmetry.
Note that the conformal mode $\psi$ is just another matter
field at the fixed point. Therefore the conformal mode
contribution can be included by the replacement
$c \rightarrow c+1$. One of the difficulties of the
$2+\epsilon$ expansion is the treatment of the conformal
mode due to the kinematical (${1\over \epsilon}$) pole in
the propagator before the reformulation of the theory.
Our calculation certainly addresses this question.
However it turns out that this problem is no more difficult
than to quantize matter fields.

In the renormalization program, the counterterms at
the $n$-th loop level are local if the theory is renormalized
at the ($n-1$)-th loop level, since all the subdivergences are
already subtracted.
Therefore the two--loop level renormalizability of the theory
at the fixed point
is already guaranteed by the previous work
\cite{AKKN}, where all the one--loop subdiagrams which appear
in the two--loop diagrams are subtracted.
In fact we show that the theory is renormalizable by adding
local counterterms in the leading order of $c$
to the two--loop level. We further demonstrate that these
counterterms can be chosen to respect the general covariance
by having the conformal anomaly vanish at the
fixed point.

\vspace{1cm}

\section{Calculation of Two--loop Counterterms}
\setcounter{equation}{0}
As is seen in the previous section, the one--loop bare action is
\begin{eqnarray}
\frac{\mu^\epsilon}{G} \int \dd^D x \sqrt{\hat{g}} \left\{
\tilde{R} L(X) + \frac{1}{2} \eta^{ij} \tilde{g}^{\mu\nu}
\partial_\mu X_i \partial_\nu X_j - \frac{AG}{\epsilon} \tilde{R} \right\},
\label{oneloopbare}
\end{eqnarray}
where $X=(\psi,\varphi_i)$ and $\eta^{ij}=\mbox{diag}(-1,1,\cdots,1)$.
Paying attention to the ultraviolet fixed point, we set
\begin{eqnarray}
L(X)=1-\frac{1}{2} \epsilon b \eta^{ij} X_i X_j,
\label{Lx}
\end{eqnarray}
where we have replaced $b$ of the previous section
with $\epsilon b /2$ to show the $\epsilon$ factor explicitly.
The  $Z_2$ symmetry of the fixed point action is preserved
also in the two--loop calculations.
As a first step towards the complete two--loop renormalization, we evaluate
only those counterterms which are proportional to the number
of matter fields in this paper.

As in the one--loop calculation, we expand the fields around the backgrounds as
$\tilde{g}_{\mu\nu}=\hat{g}_{\mu\rho}{(\ee^{h})}^\rho_{\ \nu}$, $X_i
\rightarrow \hat{X_i}+ X_i$ and employ the background gauge.
We adopt the same gauge fixing term (\ref{gaugefixing}) as in
the one--loop case,
which is not
renormalized at the one--loop level.
The ghost action (\ref{ghostaction}) is used also in this case.
In the two--loop calculations, we have to expand the action,
in general,
up to the fourth order of the quantum fields ($h_{\mu\nu}$, $ X_i$
and ghosts).
However,
since we compute the counterterms proportional to the number of matter fields,
we need only the three-- and four--point vertices
which are quadratic with respect to the $X_i$ fields.

We expand the one--loop bare action (\ref{oneloopbare}), the gauge
fixing term (\ref{gaugefixing}) and the ghost action (\ref{ghostaction})
around the background fields to a sufficient order as is explained above.
Here we exploit the formula
\beqa
\tilde{R} &=& \hat{R} -  h^\mu_{\ \nu} \hat{R}^\nu_{\ \mu} -
                                                \nabla_\mu \nabla_\nu
h^{\mu\nu}
                         +\frac{1}{4} \nabla_\rho  h^\mu_{\ \nu}
\nabla^\rho  h^\nu_{\ \mu} \nonumber \\
                & &  +\frac{1}{2} \hat{R}^\sigma_{\mu\nu\rho} h^\rho_{\ \sigma}
                                 h^{\mu\nu}
                          -\frac{1}{2}\nabla_\nu  h^\nu_{\ \mu} \nabla_\rho
                                                  h^{\rho\mu}
                          +\nabla_\mu ( h^\mu_{\ \nu} \nabla^\rho
                                                  h^\nu_{\  \rho}) + O( h^3).
\eeqa
The background metric is expanded around the flat one as
\beq
\hat{g}_{\mu\nu}=\delta_{\mu\nu}+\hat{h}_{\mu\nu},
\eeq
where $\delta^{\mu\nu} \hat{h}_{\mu\nu}=0$
can be assumed for simplicity without loss
of generality.
The propagators of the quantum fields are defined on the flat metric.
One finds that the kinetic term for the $ h_{\mu\nu}$ field is given by
$\frac{1}{4}L(\hat{X}) \nabla_\rho  h^\mu_{\ \nu}
\nabla^\rho  h^\nu_{\ \mu} $.
In order to make it canonical, we have to divide the $ h_{\mu\nu}$
field by $\sqrt{L(\hat{X})}$.
Further, we redefine the $ h_{\mu\nu}$ field as a traceless symmetric tensor
on the flat metric.
Thus we are lead to define the $ H_{\mu\nu}$ field through
\beqa
&&h^{\mu\nu}=\frac{1}{\sqrt{L(\hat{X})}}T^{\mu\nu\lambda\rho}H_{\lambda\rho},\\
&&\delta^{\mu\nu} H_{\mu\nu} =0,
\eeqa
where $T^{\mu\nu\lambda\rho}$ is defined as
\begin{eqnarray}
T^{\mu\nu\lambda\rho}=
\frac{1}{2}\left( \hat{g}^{\mu\lambda}\hat{g}^{\nu\rho}
                      +\hat{g}^{\mu\rho}\hat{g}^{\nu\lambda}
                      -\frac{2}{D}\hat{g}^{\mu\nu}\hat{g}^{\lambda\rho}\right).
\end{eqnarray}
After this prescription, we obtain the propagators and the vertices for the
$H_{\mu\nu}$, $ X_i$ and ghost fields which are required in our calculation,
as follows.

\underline{propagators}
\begin{eqnarray}
\langle H_{\mu\nu}(p) H_{\lambda\rho}(-p) \rangle = \frac{1}{p^2}
P_{\mu\nu\lambda\rho}\\
\langle X_i(p) X_j(-p) \rangle = \frac{\eta_{ij}}{p^2}\\
\langle \eta_{\mu}(p) \bar{\eta_{\nu}}(-p) \rangle
= \frac{\delta_{\mu\nu}}{p^2}
\end{eqnarray}
Here $P_{\mu\nu\lambda\rho}$ is defined as
\begin{eqnarray}
P_{\mu\nu\lambda\rho} = \delta_{\mu\lambda}\delta_{\nu\rho}
+\delta_{\mu\rho}\delta_{\nu\lambda} -\frac{2}{D} \delta_{\mu\nu}
                                                 \delta_{\lambda\rho}.
\end{eqnarray}
In the following, the index $i$ of the $X_i$--field is omitted and
$L$ is equal to $1-\frac{1}{2}\epsilon b
\hat{X}_i \hat{X}^i$.

\underline{two--point vertices}
\begin{eqnarray}
K_1^{\mu\nu\lambda\rho\alpha\beta}
\partial_\alpha H_{\mu\nu} \partial_\beta H_{\lambda\rho} &:& \nonumber\\
K_1^{\mu\nu\lambda\rho\alpha\beta}
&=& \frac{1}{4} \sqrt{\hat{g}} T^{\mu\nu\lambda\rho} \hat{g}^{\alpha\beta} -
\frac{1}{8} P^{\mu\nu\lambda\rho} \delta^{\alpha\beta}\\
K_2^{\mu\nu\lambda\rho\alpha} H_{\mu\nu} \partial_\alpha H_{\lambda\rho} &:&
\nonumber\\
K_2^{\mu\nu\lambda\rho\alpha}
&=& -\frac{i}{4L} \epsilon b \sqrt{\hat{g}} \hat{X} \partial_\beta \hat{X}
T^{\mu\nu\lambda\rho} \hat{g}^{\alpha\beta}
+ i \sqrt{\hat{g}} \hat{g}^{\alpha\beta} T^{\gamma\nu\lambda\rho}
{\hat{\Gamma}}^{\mu}_{\beta\gamma} \nonumber\\
& &- i \frac{1}{L} \epsilon b \sqrt{\hat{g}} \hat{X} \partial_\gamma \hat{X}
T_\beta^{\ \gamma\mu\nu} T^{\beta\alpha\lambda\rho}\\
K_3^{\mu\nu\lambda\rho} H_{\mu\nu}  H_{\lambda\rho} &:&
\nonumber\\
K_3^{\mu\nu\lambda\rho}
&=& \frac{1}{4L} \epsilon b \sqrt{\hat{g}} \hat{X} \partial_\gamma \hat{X}
T^{\mu\nu\alpha\rho} \hat{g}^{\beta\gamma} \hat{\Gamma}^\lambda_{\alpha\beta}
+\frac{1}{L} \epsilon b \sqrt{\hat{g}} \hat{X} \partial_\delta \hat{X}
T_\gamma^{\ \delta\mu\nu} T^{\beta\gamma\alpha\rho}
\hat{\Gamma}^\lambda_{\alpha\beta} \nonumber\\
& & +\frac{1}{4L} \epsilon b \sqrt{\hat{g}} \hat{X} \partial_\gamma \hat{X}
T^{\lambda\rho\alpha\nu} \hat{g}^{\beta\gamma} \hat{\Gamma}^\mu_{\alpha\beta}
+\frac{1}{L} \epsilon b \sqrt{\hat{g}} \hat{X} \partial_\delta \hat{X}
T_\gamma^{\ \delta\lambda\rho} T^{\beta\gamma\alpha\nu}
\hat{\Gamma}^\mu_{\alpha\beta} \nonumber\\
& & -\sqrt{\hat{g}} T^{\alpha\nu\delta\rho} \hat{g}^{\beta\gamma}
\hat{\Gamma}^\mu_{\alpha\beta}
\hat{\Gamma}^\lambda_{\gamma\delta}
-\frac{1}{2} \sqrt{\hat{g}} \hat{R}^\alpha_{\beta\gamma\delta}
T_\alpha^{\ \delta\mu\nu} T^{\beta\gamma\lambda\rho} \nonumber\\
& & -\frac{1}{4L} \sqrt{\hat{g}} \partial_\alpha\hat{X} \partial_\beta\hat{X}
T_\gamma^{\ \alpha\mu\nu} T^{\beta\gamma\lambda\rho}\\
K_4^{\mu\nu\alpha}  H_{\mu\nu} \partial_\alpha  X  &:&
\nonumber\\
K_4^{\mu\nu\alpha}
&=& i \frac{1}{\sqrt{L}} \sqrt{\hat{g}} \partial_\beta \hat{X}
T^{\alpha\beta\mu\nu}\\
K_5^{\mu\nu} H_{\mu\nu}  X  &:&
\nonumber\\
K_5^{\mu\nu}
&=& -\frac{1}{\sqrt{L}} \epsilon b \sqrt{\hat{g}} \hat{X}
\hat{R}_{\alpha\beta} T^{\alpha\beta\mu\nu}\\
K_6^{\alpha\beta} \partial_\alpha  X \partial_\beta   X &:&
\nonumber\\
K_6^{\alpha\beta}
&=& \frac{1}{2} (\sqrt{\hat{g}}\hat{g}^{\alpha\beta}-\delta^{\alpha\beta})\\
K_7  X^2 &:& \nonumber\\
K_7
&=& \frac{1}{2} \epsilon b \sqrt{\hat{g}} \hat{R}\\
\tilde{K}_1^{\mu\nu\lambda\rho\alpha\beta}
\partial_\alpha H_{\mu\nu} \partial_\beta H_{\lambda\rho} &:&
\nonumber\\
\tilde{K}_1^{\mu\nu\lambda\rho\alpha\beta}
&=& -\frac{AG}{\epsilon}\left(\frac{1}{4L} \sqrt{\hat{g}}
T^{\mu\nu\lambda\rho} \hat{g}^{\alpha\beta}
-\frac{1}{2L} \sqrt{\hat{g}}
T_\gamma^{\ \alpha\mu\nu} T^{\gamma\beta\lambda\rho}\right)\\
\tilde{K}_2^{\mu\nu\lambda\rho\alpha}
H_{\mu\nu} \partial_\alpha H_{\lambda\rho} &:&
\nonumber\\
\tilde{K}_2^{\mu\nu\lambda\rho\alpha}
&=& -\frac{AG}{\epsilon}\left(-\frac{i}{4L^2} \epsilon b \sqrt{\hat{g}}
\hat{X} \partial_\beta\hat{X}
T^{\mu\nu\lambda\rho} \hat{g}^{\alpha\beta}
+i \frac{1}{L} \sqrt{\hat{g}} T^{\gamma\nu\lambda\rho}
\hat{g}^{\alpha\beta} \hat{\Gamma}^\mu_{\beta\gamma} \right.\nonumber\\
& & \left. -i \frac{2}{L} \sqrt{\hat{g}}
T^{\gamma\beta\delta\nu} T_\beta^{\ \alpha\lambda\rho}
\hat{\Gamma}^\mu_{\gamma\delta}
+\frac{i}{2L^2} \epsilon b \sqrt{\hat{g}} \hat{X} \partial_\beta \hat{X}
T_\gamma^{\ \beta\mu\nu} T^{\alpha\gamma\lambda\rho} \right)\\
\tilde{K}_3^{\mu\nu\lambda\rho} H_{\mu\nu}  H_{\lambda\rho} &:&
\nonumber\\
\tilde{K}_3^{\mu\nu\lambda\rho}
&=& -\frac{AG}{\epsilon}\left(\frac{1}{4L^2} \epsilon b \sqrt{\hat{g}}
\hat{X} \partial_\gamma \hat{X} T^{\mu\nu\alpha\rho}
\hat{g}^{\beta\gamma} \hat{\Gamma}^\lambda_{\alpha\beta}
-\frac{1}{2 L^2} \epsilon b \sqrt{\hat{g}} \hat{X} \partial_\delta \hat{X}
T_\gamma^{\ \delta\mu\nu} T^{\gamma\beta\alpha\rho}
\hat{\Gamma}^\lambda_{\alpha\beta} \right. \nonumber\\
& & +\frac{1}{4 L^2} \epsilon b \sqrt{\hat{g}} \hat{X} \partial_\gamma \hat{X}
T^{\lambda\rho\alpha\nu} \hat{g}^{\beta\gamma} \hat{\Gamma}^\mu_{\alpha\beta}
-\frac{1}{2 L^2} \epsilon b \sqrt{\hat{g}} \hat{X} \partial_\delta \hat{X}
T_\gamma^{\ \delta\lambda\rho} T^{\gamma\beta\alpha\nu}
\hat{\Gamma}^\mu_{\alpha\beta} \nonumber\\
& & -\frac{1}{L} \sqrt{\hat{g}} T^{\alpha\nu\delta\rho} \hat{g}^{\beta\gamma}
\hat{\Gamma}^\mu_{\alpha\beta} \hat{\Gamma}^\lambda_{\gamma\delta}
-\frac{1}{2L} \sqrt{\hat{g}} \hat{R}^\alpha_{\beta\gamma\delta}
T_\alpha^{\ \delta\mu\nu} T^{\beta\gamma\lambda\rho} \nonumber\\
& & \left. +\frac{2}{L} \sqrt{\hat{g}}
T_\alpha^{\ \beta\gamma\nu} T^{\delta\alpha\eta\rho}
\hat{\Gamma}^\mu_{\beta\gamma} \hat{\Gamma}^{\lambda}_{\delta\eta}\right) \\
\bar{K}_1^{\mu\nu\alpha\beta}
\partial_\alpha \bar{\eta}_\mu  \partial_\beta \eta_\nu &:&
\nonumber\\
\bar{K}_1^{\mu\nu\alpha\beta}
&=& \sqrt{\hat{g}} \hat{g}^{\mu\nu} \hat{g}^{\alpha\beta}
-\delta^{\mu\nu} \delta^{\alpha\beta}\\
\bar{K}_2^{\mu\nu\alpha} \partial_\alpha \bar{\eta}_\mu \eta_\nu &:&
\nonumber\\
\bar{K}_2^{\mu\nu\alpha}
&=& i \sqrt{\hat{g}} \hat{g}^{\alpha\beta} \hat{g}^{\mu\gamma}
\hat{\Gamma}^\nu_{\beta\gamma}\\
\bar{K}_3^{\mu\nu\alpha}  \bar{\eta}_\mu \partial_\alpha \eta_\nu &:&
\nonumber\\
\bar{K}_3^{\mu\nu\alpha}
&=& i \sqrt{\hat{g}} \hat{g}^{\alpha\beta} \hat{g}^{\nu\gamma}
\hat{\Gamma}^\mu_{\beta\gamma}\\
\bar{K}_4^{\mu\nu}  \bar{\eta}_\mu \eta_\nu &:&
\nonumber\\
\bar{K}_4^{\mu\nu}
&=& -\sqrt{\hat{g}} \hat{g}^{\alpha\beta} \hat{g}^{\gamma\delta}
\hat{\Gamma}^\mu_{\alpha\gamma} \hat{\Gamma}^\nu_{\beta\delta}
-\sqrt{\hat{g}} \hat{R}^{\mu\nu}
\end{eqnarray}
\underline{three--point vertices}
\begin{eqnarray}
V_1^{\mu\nu\alpha\beta} H_{\mu\nu} \partial_\alpha  X \partial_\beta  X &:&
\nonumber\\
V_1^{\mu\nu\alpha\beta}
&=& -\frac{1}{2\sqrt{L}} \sqrt{\hat{g}}
T^{\mu\nu\alpha\beta}\\
V_{2}^{\mu\nu} H_{\mu\nu}  X^2 &:&
\nonumber\\
V_{2}^{\mu\nu}
&=& -\frac{1}{2\sqrt{L}} \epsilon b \sqrt{\hat{g}} \hat{R}_{\alpha\beta}
T^{\mu\nu\alpha\beta}
\end{eqnarray}
\underline{four--point vertices}
\begin{eqnarray}
W_1^{\mu\nu\lambda\rho\alpha\beta}
\partial_\alpha H_{\mu\nu} \partial_\beta H_{\lambda\rho}  X^2 &:&
\nonumber\\
W_1^{\mu\nu\lambda\rho\alpha\beta}
&=& -\frac{1}{8L}\epsilon b \sqrt{\hat{g}}
T^{\mu\nu\lambda\rho} \hat{g}^{\alpha\beta}\\
W_2^{\mu\nu\lambda\rho\alpha\beta}
H_{\mu\nu} \partial_\alpha H_{\lambda\rho}  X \partial_\beta  X &:&
\nonumber\\
W_2^{\mu\nu\lambda\rho\alpha\beta}
&=& \frac{1}{L} \epsilon b \sqrt{\hat{g}}
T_\gamma^{\ \beta\mu\nu} T^{\gamma\alpha\lambda\rho}\\
W_{3}^{\mu\nu\lambda\rho\alpha\beta}
H_{\mu\nu}  H_{\lambda\rho} \partial_\alpha  X  \partial_\beta  X &:&
\nonumber\\
W_{3}^{\mu\nu\lambda\rho\alpha\beta}
&=& \frac{1}{8L} \sqrt{\hat{g}}
(T_\gamma^{\ \alpha\mu\nu} T^{\gamma\beta\lambda\rho}
+ T_\gamma^{\ \beta\mu\nu} T^{\gamma\alpha\lambda\rho})\\
W_{4}^{\mu\nu\lambda\rho\alpha}
H_{\mu\nu} \partial_\alpha H_{\lambda\rho}  X^2 &:&
\nonumber\\
W_{4}^{\mu\nu\lambda\rho\alpha}
&=& -i \frac{1}{2L} \epsilon b \sqrt{\hat{g}}
T^{\beta\nu\lambda\rho} \hat{g}^{\alpha\gamma}
\hat{\Gamma}^\mu_{\beta\gamma}\\
W_{5}^{\mu\nu\lambda\rho\alpha}
H_{\mu\nu} H_{\lambda\rho}  X \partial_\alpha  X &:&
\nonumber\\
W_{5}^{\mu\nu\lambda\rho\alpha}
&=& i\frac{1}{L} \epsilon b \sqrt{\hat{g}}
T_\delta^{\ \alpha\mu\nu} T^{\delta\beta\gamma\rho}
\hat{\Gamma}^\lambda_{\beta\gamma} \nonumber\\
& &+i\frac{1}{L} \epsilon b \sqrt{\hat{g}}
T_\delta^{\ \alpha\lambda\rho} T^{\delta\beta\gamma\nu}
\hat{\Gamma}^\mu_{\beta\gamma}\\
W_{6}^{\mu\nu\lambda\rho}
H_{\mu\nu} H_{\lambda\rho}  X^2 &:&
\nonumber\\
W_{6}^{\mu\nu\lambda\rho}
&=& \frac{1}{2L} \epsilon b \sqrt{\hat{g}}
T^{\alpha\nu\delta\rho} \hat{g}^{\beta\gamma}
\hat{\Gamma}^\mu_{\alpha\beta} \hat{\Gamma}^\lambda_{\gamma\delta} \nonumber\\
& &+\frac{1}{4L} \epsilon b \sqrt{\hat{g}}
\hat{R}^\alpha_{\beta\gamma\delta}
T_\alpha^{\ \delta\mu\nu} T^{\beta\gamma\lambda\rho}\\
\bar{W}_1^{\mu\nu\alpha\beta}
\partial_\alpha \bar{\eta}_\mu \eta_\nu  X \partial_\beta  X &:&
\nonumber\\
\bar{W}_1^{\mu\nu\alpha\beta}
&=& \epsilon b \sqrt{\hat{g}} \hat{g}^{\mu\alpha} \hat{g}^{\nu\beta}\\
\bar{W}_2^{\mu\nu\alpha}
\bar{\eta}_\mu \eta_\nu  X \partial_\alpha   X &:&
\nonumber\\
\bar{W}_2^{\mu\nu\alpha}
&=& -\epsilon b \sqrt{\hat{g}} \hat{g}^{\beta\gamma} \hat{g}^{\nu\alpha}
\hat{\Gamma}^\mu_{\beta\gamma}
\end{eqnarray}

In the forthcoming figures which represent the Feynman diagrams,
the wavy lines and the solid lines represent
the propagators of the $H_{\mu\nu}$ and $ X_i$ fields respectively.
The dots denote the derivatives and
the circle with the vertical line represents the one--loop
counterterm insertion.

To determine the counterterms, we make use of the manifest general
covariance with respect to the background metric $\hat{g}_{\mu\nu}$.
We keep the contributions of the appropriate order in $\hat{h}_{\mu\nu}$
in the calculation
of the diagrams and read off the general covariant forms from the results.

Our strategy of the calculation is as follows.
Firstly, we set $\hat{h}_{\mu\nu}=0$ and evaluate the divergences proportional
to $\partial_\mu \hat{X}_i \partial_\mu \hat{X}^i/2$ to determine
the counterterm for
$\tilde{g}^{\mu\nu}\partial_\mu X_i \partial_\nu X^i/2$.
Secondly, we calculate the diagrams
which are of the first order in $\hat{h}_{\mu\nu}$. We subtract
from the results
the $O(\hat{h})$ contributions coming from
the term $\hat{g}^{\mu\nu} \partial_\mu \hat{X}_i \partial_\nu \hat{X}^i/2$
which are derived in the first step. Exploiting
$\sqrt{\hat{g}}\hat{R}=-\partial_\mu \partial_\nu \hat{h}_{\mu\nu}
+O(\hat{h}^2)$, we obtain
the counterterm for the $X$--dependent part of $\tilde{R}L(X)$.
Finally, we compute the diagrams which are of the second order in
$\hat{h}_{\mu\nu}$ after setting $\hat{X}_i=0$.
By making use of the relation
\begin{eqnarray}
\int \dd^D x \sqrt{\hat{g}} \hat{R} = \int \dd^D x
\left(\frac{1}{4}\partial_\mu\hat{h}_{\lambda\rho}
\partial_\mu\hat{h}_{\lambda\rho}
-\frac{1}{2}
\partial_{\mu}\hat{h}_{\mu\nu}\partial_\rho\hat{h}_{\rho\nu}\right)
+O(\hat{h}^3),
\label{Rhat}
\end{eqnarray}
we fix the counterterm for $\tilde{R}$.

In order to renormalize the theory up to the two--loop level,
we have to make sure that the two--loop divergences are local.
It is possible to do so only if we subtract the subdivergences
of the one--loop subdiagrams in the two--loop diagrams properly.
The one--loop renormalization of the quantum fields shows
that the only one--loop counterterm proportional
to $c$ is ${c \over 24\pi \epsilon} \tilde R$ .
This means that the subdivergences should arise only from the matter
subloops connected to the quantum $H_{\mu\nu}$
or the background $\hat h_{\mu\nu}$ lines.
Keeping this point in mind, we can classify all the diagrams
into groups, each of which gives local divergences.

Also we comment on a subtlety in calculating
the short distance divergences in two--loop diagrams;
namely the subdiagrams containing the $ H_{\mu\nu}$ propagators,
in general, cause infrared divergences.
In order to regularize them, we introduce a  mass term in the $H_{\mu\nu}$
propagator as $\frac{1}{p^2} \rightarrow \frac{1}{p^2 + m^2}$. We take
the $m \rightarrow  0$ limit after extracting the
$\frac{1}{\epsilon}\log (\frac{m^2}{k^2})$
type divergences.
As is seen later, such divergences cancel out among the diagrams and
do not appear in the final results. Therefore the short distance divergences
are
separated from the infrared divergences and there are no mixed divergences.

\vspace{1cm}

\section{Results for Two--loop Counterterms}
\setcounter{equation}{0}
In this section, we calculate the
two--loop divergences in the effective
action following the strategy described in the previous section
and show the results in detail.
The two--loop counterterms are readily obtained by the replacements
$\hat{X}_i \rightarrow X_i$ and
$\hat{g}_{\mu\nu} \rightarrow \tilde{g}_{\mu\nu}$ in the
two--loop divergences.

\subsection{Divergences for $\hat{g}^{\mu\nu} \partial_\mu \hat{X}_i
\partial_\nu \hat{X}^i$}
In order to evaluate the divergences for the kinetic term of $\hat{X}_i$ ,
we have to consider six diagrams (Fig.1), where  $\hat{h}_{\mu\nu}$ is set to
be
equal to zero.
The results are
obtained in the form
\beq
\frac{cG}{(4\pi)^2}  \alpha \int \dd^D x \frac{1}{2} \partial_\mu \hat{X}_i
\partial_\mu \hat{X}^i,
\eeq
and the coefficient $\alpha$ for each diagram is shown
in Table 1. We can see that
each of the diagrams gives a local single--pole divergence
and has no infrared divergence.
The final result is written in the covariant form as
\beq
\frac{cG}{(4\pi)^2}  \biggl( -{1 \over 8 \epsilon} \biggr)
\int \dd^D x \sqrt{\hat{g}}
\hat{g}^{\mu\nu} \frac{1}{2} \partial_\mu \hat{X}_i \partial_\nu \hat{X}^i.
\eeq

\subsection{Divergences for $\hat{R} \  \hat{X}_i \hat{X}^i$}
We write down all the diagrams which are
of the first order in $\hat{h}_{\mu\nu}$ and
subtract
from them the contributions of
$\hat{g}^{\mu\nu} \partial_\mu \hat{X}_i \partial_\nu \hat{X}^i$.
Thus we obtain thirty--three diagrams, which is found to
give the divergences of the form $\hat R \ \hat X_i \hat X^i$.
Note here that since all the diagrams include a vertex proportional
to $\epsilon  b$,
there is in principle neither nonlocal nor infrared divergence,
though one may have a local single--pole divergence in general.
We can, therefore, conclude that
the theory is renormalizable with $L=1-{1\over 2}\epsilon b X_iX^i$
up to the two--loop level, though arbitrary polynomials of $X_iX^i$
may appear in $L$ at higher orders.
A concrete calculation shows that
only five diagrams provide nontrivial contributions,
which cancel among them.
Thus, as a result, we find no divergences for $\hat R \ \hat X_i \hat X^i$.

\subsection{Divergences for $ \hat{R}$}
We set $\hat{X}_i=0$ and evaluate the two--point functions of
$\hat{h}_{\mu\nu}$.
The diagrams we have to calculate are classified into two categories.
The first one consists of forty--one diagrams, which contain a vertex
proportional to
$\epsilon b$ and give only a local single--pole divergence in general.
Among them, there are
four diagrams with the ghost loop and each of them is found
to give no contribution.
As for the remaining thirty--seven diagrams,
our calculation shows
that all the nonvanishing contributions cancel among the diagrams.

The other category is a set of forty--three diagrams which possess no overall
$\epsilon b$ factor and
are able to give nonlocal as well as infrared divergences.
We classify them into the thirteen groups (Fig.2--Fig.14), such that
the contribution from each of them becomes local.

Groups $1,\ 2$ and $3$, as well as Group 5 to 11,
are topologically the same,
but differ in the vertices connected to the external $\hat{h}_{\mu\nu}$ lines.
Except for Group 13,
the one--loop counterterm
insertion cancels the subdivergence from the matter subloop.
On the other hand, the two diagrams of Group $13$ contain
the subdiagrams which are the two--point functions of the matter fields
at the one--loop level.
As is seen in the one--loop calculation,
the divergent contributions from these subdiagrams cancel each other,
which implies that Group $13$ gives a local divergence without
one--loop counterterm insertions.
The calculations of the diagrams with the same topology as the diagram
$1-1$ are carried out by the method presented in ref. \cite{IZ}.

We summarize the results for each of the Groups in Tables $2$ to $14$.
We obtain  generally the divergences in the momentum space as
\beqa
&~& \frac{cG}{(4\pi)^2}\int \frac{\dd^D k}{{(2\pi)}^D} \biggl\{
A k_\mu k _\nu \hat{h}(k)_{\mu\lambda}
\hat{h}(-k)_{\nu\lambda}+B k^2 \hat{h}(k)_{\mu\nu} \hat{h}(-k)_{\mu\nu}
\nonumber \\
&&~~~~~~~~~~~~~~~~~~~~~+C\frac{1}{k^2}k_\mu k_\nu \hat{h}(k)_{\mu\nu}
k_\lambda k_\rho \hat{h}(-k)_{\lambda\rho} \biggr\}.
\eeqa
The coefficients $A$, $B$ and $C$ are shown in the Tables. The
$\rho$ and $\sigma$ in the Tables are defined as,
\beqa
\rho= \log \left(\frac{k^2}{4\pi}\right),\nonumber \\
\sigma=\log \left(\frac{m^2}{k^2}\right). \nonumber
\eeqa
We can see that in the total of each of the
Groups, $\rho$ and $\sigma$ do not appear and $C$ is equal to zero .
This means that the nonlocal divergences as well as
the infrared divergences in the $\frac{1}{\epsilon}$ pole cancel
among the diagrams within each of the Groups.
We also collect the total results of the Groups in
Table $15$ and sum them up in the total.
The double--pole singularity does vanish in the final result
although it remains in each Group generically.
 From the relation between $A$ and $B$ in the final result
and the formula (\ref{Rhat}),
we can verify the preservation of the general covariance
with respect to the background,
which serves as a check of our calculation.
The final result for the divergent contribution is found to be
\beq
\frac{cG}{{(4\pi)}^2} \frac{5}{24\epsilon} \int \dd^D x \sqrt{\hat{g}} \hat{R}.
\eeq


\vspace{1cm}

\section{Conformal Invariance and the Ultraviolet Fixed Point}
\setcounter{equation}{0}
In the previous section, we have calculated the divergences in the effective
action and determined the counterterms by the background field method.
The bare action can be written as
\beqa
S_0 &=& \int \dd^D x  \sqrt{\hat{g}} \frac{\mu^{\epsilon}}{G}
        \left\{ \left( 1 - \frac{25-c}{6 \epsilon} \hat{G} +
             \frac{5c}{24 \epsilon} \hat{G}^2 \right) \tilde{R} \right.\n
    && ~~~~~~~~~~~~~~~~~~~  - \frac{1}{2} ( \epsilon b - w \hat{G}^2 b)
           X_i X^i \tilde{R} \n
    && ~~~~~~~~~~~~~~~~~~~ \left. + \frac{1}{2} \left( 1- \frac{c}{8 \epsilon}
\hat{G}^2 \right)
         \tilde{g} ^{\mu\nu} \del_{\mu} X_i \del_{\nu} X^i \right\},
\eeqa
where $\hat{G}$ is equal to $G/4\pi$.
We have introduced a parameter $w$ , which corresponds to a finite
renormalization of the coupling $X_i X^i \tilde{R}$.
Although it can be taken arbitrary as far as the divergence of the theory is
concerned, we have to keep it since the corresponding tree--level coupling
constant is $O(\epsilon)$.

We parametrize the bare action as
\beq
S_0 = \frac{1}{G_0} \int \dd^D x  \sqrt{\hat{g}} \left\{
\left(1-\frac{1}{2} \epsilon b_0 X_{0i} X_0^{~i}\right) \tilde{R}
+ \frac{1}{2} \tilde{g}^{\mu\nu}\del_{\mu} X_{0i} \del_{\nu} X_0^{~i} \right\},
\eeq
which gives the following relations between the bare quantities
and the renormalized quantities.
\beqa
\frac{1}{G_0} &=& \frac{\mu^{\epsilon}}{G}
                \left( 1 - \frac{25-c}{6 \epsilon} \hat{G}
                 + \frac{5c}{24\epsilon} \hat{G}^2 \right) , \label{eq:g0}\\
\epsilon b_0 &=& \epsilon b + \left( \frac{c}{8} - w \right) \hat{G}^2 b.
\label{eq:b0}
\eeqa

Using these relations, the $\beta$ functions can be obtained as
\beqa
\beta_G &=& G \left( \epsilon - \frac{25-c}{6} \hat{G}
            + \frac{5c}{12} \hat{G}^2 \right) , \\
\beta_b &=& \left( 2w - \frac{c}{4} \right) \hat{G}^2 b.
\eeqa
The absence of the double pole at the two--loop level in (\ref{eq:g0})
follows from the finiteness of $\beta_G$ just like in the Yang-Mills
theory with the background gauge \cite{Abbott}.
The absence of the
single pole in (\ref{eq:b0}) is also required for the the finiteness
of $\beta_b$. This coupling constant always has an $\epsilon$ factor
suppression due to the symmetry under the constant shift of the $X_i$
fields.
One finds, in the expression for $\beta_b$, that the free parameter $w$ is
relevant to the physics of the system.
As we see in the following, this ambiguity will be fixed
when we impose the general covariance on the bare action.

Since we have maintained only
the volume--preserving diffeomorphism, we have to
impose the conformal invariance on the bare action so that the theory is
generally covariant.
We consider the conformal transformation
\beqa
\delta \tilde{g}_{\mu\nu} &=& \tilde{g}_{\mu\nu} \delta\rho , \\
\delta X_{0i} &=& (D-1) \frac{\del L_0}{\del X_0^{~i}} \delta \rho ,
\eeqa
where $L_0=1-\frac{1}{2} \epsilon b_0 X_{0 i} X_0^{~i}$.
Under this transformation,
the bare action transforms as
\beqa
\delta S_0 &=&
\int \dd^D x  \sqrt{\hat{g}} \frac{\mu^{\epsilon}}{2G} \left[
\left\{ \epsilon - \frac{25-c}{6} \hat{G} + \frac{5c}{24} \hat{G}^2
\right\} \tilde{R} \right.\n
&&- \frac{1}{2} \left\{ \epsilon ( \epsilon b - w \hat{G}^2 b )
            -4 (D-1) ( \epsilon b - w \hat{G}^2 b)
       \left( \epsilon - w \hat{G}^2 + \frac{c}{8} \hat{G}^2\right)b \right\}
       X_i X^i \tilde{R} \n
&&+ \left. \frac{1}{2} \left\{
  \epsilon \left( 1 - \frac{c}{8 \epsilon} \hat{G}^2 \right)
  - 4(D-1)(\epsilon b - w \hat{G}^2 b) \right\}
      \del_{\mu} X_i \del^{\mu} X^i \right]\delta \rho.
\label{eq:anomaly}
\eeqa
When we consider symmetry at the quantum level within the counterterm
formalism, we have to replace the operators in (\ref{eq:anomaly})
with the corresponding renormalized operators
\cite{CFMP,FT,Tseytlin}.

In order to define the renormalized operator for $X_i X^i \tilde{R}$,
we differentiate the bare action with respect to the finite parameter $b$
\beq
\int \dd^D x  \sqrt{\hat{g}} ( X_i X^i \tilde{R} )_r \equiv
 - \frac{2G}{\epsilon \mu^{\epsilon}} \frac{\del S_0}{\del b}
= \left( 1 - \frac{w \hat{G}^2}{\epsilon} \right) \int \dd^D x
\sqrt{\hat{g}}  X_i X^i \tilde{R} .
\label{eq:preXXtildeRren}
\eeq
We need to translate this relation into the local one,
where, in general, one may have some total derivative terms.
We note, however, that a complete set of operators can be written without
total derivative terms by making use of the equations of motion
\cite{Tseytlin}.
We can, therefore, define the renormalized operator
$( X_i X^i \tilde{R} )_r$ as
\beq
 ( X_i X^i \tilde{R} )_r
= \left( 1 - \frac{w \hat{G}^2}{\epsilon} \right)   X_i X^i \tilde{R} .
\label{eq:XXtildeRren}
\eeq
The same reasoning holds in the case of the other operators, which
we omit to mention in the following.

The renormalized operator for $\del_{\mu} X_i \del^{\mu} X^i$ can be
obtained by introducing a parameter ``$f$'' in front of the tree--level
kinetic term of $X_i$ and keep track of the parameter in the divergent
diagrams.
The propagator of $X_i$ is multiplied by a factor $\frac{1}{f}$ and the
vertices which originate from the kinetic term of $X_i$ are multiplied by
a factor $f$.
One finds that all the diagrams corresponding to the renormalization of the
kinetic term are multiplied by a factor $f$ and the other diagrams remain
the same.
This implies that the renormalized operator for
$\del_{\mu} X_i \del^{\mu} X^i$ can be defined through
\beq
\del_{\mu} X_i \del^{\mu} X^i =
\left( 1 + \frac{c\hat{G}^2}{8 \epsilon} \right)
(\del_{\mu} X_i \del^{\mu} X^i )_r.
\label{eq:delXdelXren}
\eeq

Finally the renormalized operator for $\tilde{R}$ can be obtained as follows.
We renormalize $X_i$ so that the only $G$ dependence comes from the
coefficient of $\tilde{R}$
\footnote{Here we assume that
the $G$ dependence in the gauge fixing term does not
affect the physical conclusion. An explicit check of this assumption
by performing the operator renormalization of $\tilde{R}$ requires as much
work as has been done in this study.}.
Also we have to perform the wave function renormalization
in order to avoid picking
up unphysical contributions from the kinetic term of $X_i$.
Thus we define
\beq
Y_i = \sqrt{ \frac{1}{G} \left(1-\frac{c}{8 \epsilon} \hat{G}^2 \right)} X_i ,
\eeq
and rewrite the action in terms of $Y_i$ as
\beqa
S_0(G,Y_i) &=& \int \dd^D x  \sqrt{\hat{g}}  \mu^\epsilon \left[ \frac{1}{G}
    \left( 1 - \frac{25-c}{6\epsilon} \hat{G}
   + \frac{5c}{24\epsilon}\hat{G}^2 \right) \tilde{R} \right.\n
 && ~~~~~~~~~~~~~~~~~- \frac{1}{2} \left( \epsilon b - w \hat{G}^2 b +
               \frac{c}{8} \hat{G}^2 b
      \right)  Y_i Y^i \tilde{R} \n
 && ~~~~~~~~~~~~~~~~~\left. +\frac{1}{2} \del_{\mu} Y_i \del^{\mu} Y^i \right].
\eeqa
By differentiating the above bare action with respect to $1/G$, we can obtain
the renormalized operator for $\tilde{R}$ as
\beq
(\tilde{R})_r
= \left( 1 - \frac{5c}{24 \epsilon} \hat{G}^2 \right) \tilde{R}
  - \left( w \hat{G}^2 b - \frac{c}{8} \hat{G}^2 b \right) X_i X^i \tilde{R} .
\label{eq:tildeRren}
\eeq
Using eqs. (\ref{eq:XXtildeRren}), (\ref{eq:delXdelXren}) and
(\ref{eq:tildeRren}), the conformal anomaly (\ref{eq:anomaly})
 can be written in terms of the renormalized operators as
\beqa
&&  \int \dd^D x  \sqrt{\hat{g}} \frac{\mu^{\epsilon}}{2G}
\left[ \left\{ \epsilon - \frac{25-c}{6}\hat{G}
+ \frac{5c}{12} \hat{G}^2 \right\}   ( \tilde{R})_r \right. \n
&& ~~~~~~~~~ - \frac{1}{2} \epsilon b
 \left\{ \epsilon - 4 (D-1)
  \left( \epsilon b - w \hat{G}^2 b + \frac{c}{8} \hat{G}^2 b \right)
-2  \left(w \hat{G}^2  - \frac{c}{8} \hat{G}^2 \right) \right\}
 ( X_i X^i \tilde{R} )_r \n
&& ~~~~~~~~~ - \left. \frac{1}{2}
  \left\{ \epsilon - 4(D-1)\left(
    \epsilon b - w \hat{G}^2 b + \frac{c}{8} \hat{G}^2 b \right) \right\}
   (\del_{\mu} X_i \del^{\mu} X^i )_r \right] \delta \rho.
\eeqa

This result is reasonable since each term includes the expression for the
$\beta$ function.
At the ultraviolet fixed point, where the $\beta$ functions vanish,
the conformal anomaly vanishes if and only if the fixed--point value of $b$ is
given by
\beq
b= b^{\ast} = \frac{1}{4(D-1)}.
\eeq
This is the same as the one--loop result.
In order that this nonvanishing fixed--point value of $b$ may be realized,
the coefficient $(2w- c/4)$ in the $\beta$ function of $b$ should vanish.
Thus the free parameter $w$ should be chosen to be $\frac{c}{8}$
to the leading order of $c$.
Note also that the fixed--point value of $b$ coincides with
the value of $b$ that corresponds to the classical Einstein gravity.
This is consistent with the one--loop result where it has been shown that
the $\beta$ function of $b$ remains zero
throughout the renormalization group trajectory
from the ultraviolet fixed point to the infrared fixed point which
corresponds to Einstein gravity.

\vspace{1cm}

\section{Summary and Discussion}
\setcounter{equation}{0}
The recent progress in two-dimensional quantum gravity
has provided us with an example of a consistent field theory with the
general covariance.
It is natural for us to hope that we can formulate quantum gravity near
two dimensions by using the $\epsilon$ expansion.
It has been discovered,
however, that special care should be taken when we impose
the general covariance on the theory in the renormalization procedure.
It is because the general covariance inevitably relates the large scale physics
to the short distance physics. This feature is not easy to reconcile
with the idea of the renormalized field theory, where we consider the physics
at a fixed scale.

Putting it in a slightly different way,
the subtlety arises from the fact that in quantum gravity
we integrate over the metric which serves to set the physical scale.
We, therefore, have to consider all length scale at once.
However in field theory, we need to introduce the short distance cutoff
in some form, which inevitably breaks the general covariance.
The conformal anomaly may be a manifestation of such difficulties.
The oversubtraction problem \cite{KKN1,KKN2} and the conformal anomaly
are the different faces of the same coin.
In this sense we are dealing with a generic problem in quantum gravity
which is not specific to the Einstein gravity near two dimensions.

The strategy adopted in ref. \cite{AKKN} was to take the tree action to
be the most general one that is invariant under the volume--preserving
diffeomorphism and to impose the full general covariance by
choosing a renormalization group
trajectory on which the conformal anomaly vanishes.
At the one--loop level it has been shown that there exists such a trajectory
which starts from an ultraviolet fixed point and flows into the
classical Einstein gravity in the infrared limit.
However, it is certainly nontrivial whether this idea works to all orders of
the $\epsilon$ expansion.
It is, therefore, desirable to see how it goes at the two--loop level.
The results we have obtained in this paper are very encouraging.
Although the conformal anomaly inevitably arises in quantum field theory, it is
a short distance effect and always local. Therefore we should be able to
cancel it by changing the coupling constants in the theory.
Based on these reasonings,
we expect that this program will succeed.

In this paper, we have studied two--loop renormalization imposing the $Z_2$
symmetry on the system.
We have calculated the divergences proportional to the number of matter fields
and examined how the nonlocal divergences
as well as the infrared divergences in the
$\frac{1}{\epsilon}$
pole cancel
among the diagrams.
It has been shown that the conformal anomaly vanishes at the ultraviolet
fixed point when we choose the finite renormalization properly.
This ensures the existence of the ultraviolet fixed point which possesses
the general covariance up to the two--loop level in the leading order of $c$.
We have to work out similar calculations without imposing the $Z_2$ symmetry
on the system in order to examine the general covariance on the
renormalization group trajectory that flows into the classical Einstein
gravity.
It seems also important for us to perform the full calculation of
the two--loop renormalization without restricting ourselves to the leading
matter contribution.
We hope that we can eventually calculate physical quantities
such as the critical exponents,
which may be calculable also in numerical simulations of
three or four dimensional quantum gravity in future.

\vspace{1cm}

We would like to thank H. Kawai and M. Ninomiya for stimulating discussion.
Tremendous amount of tensor calculation involved in this work
has been done with the aid of MathTensor,
a software designed for symbolic manipulations in tensor analysis.
It is our pleasure to acknowledge S. Christensen of MathSolution Inc.
for his kind advice concerning the usage of this powerful tool.

\newpage
\setlength{\baselineskip}{7mm}

\newpage

\begin{figure}
 \leavevmode
 \epsfxsize = 10 cm
 \centerline{ \epsfbox{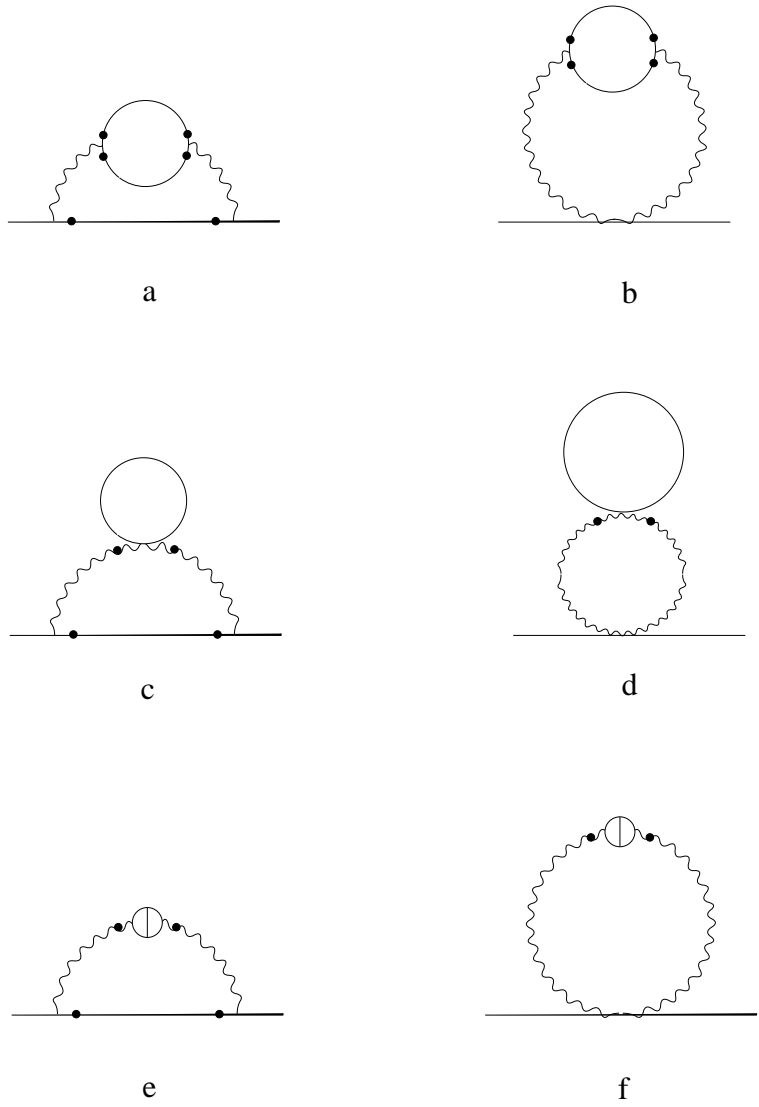} }
 \caption{ Diagrams \ for \ $\partial_\mu \hat X_i \partial_\mu \hat X^i$ }
 \label{}
\end{figure}

\begin{table}
\begin{center}
 \doublerulesep=0pt
 \def\arraystretch{1.7}
 \begin{tabular}{|c|c|c|c|c|c|c|c|}  \hline\hline
Diagram & $a$ & $b$ & $c$ & $d$ & $e$ & $f$ & Total \\  \hline
 $\alpha$  & $0 $ & $\frac{1}{8\epsilon}$ &$\frac{2b}{\epsilon}$
  & $-\frac{2b}{\epsilon}$
 &$-\frac{1}{12\epsilon}$  & $-\frac{1}{6\epsilon}$ &
$-\frac{1}{8\epsilon}$ \\  \hline\hline
 \end{tabular}
\end{center}
\caption{ The divergences for $\partial_\mu \hat X_i \partial_\mu \hat X^i$}
\label{matterkin}
\end{table}%

\begin{figure}
 \leavevmode
 \epsfxsize = 10 cm
 \centerline{ \epsfbox{ 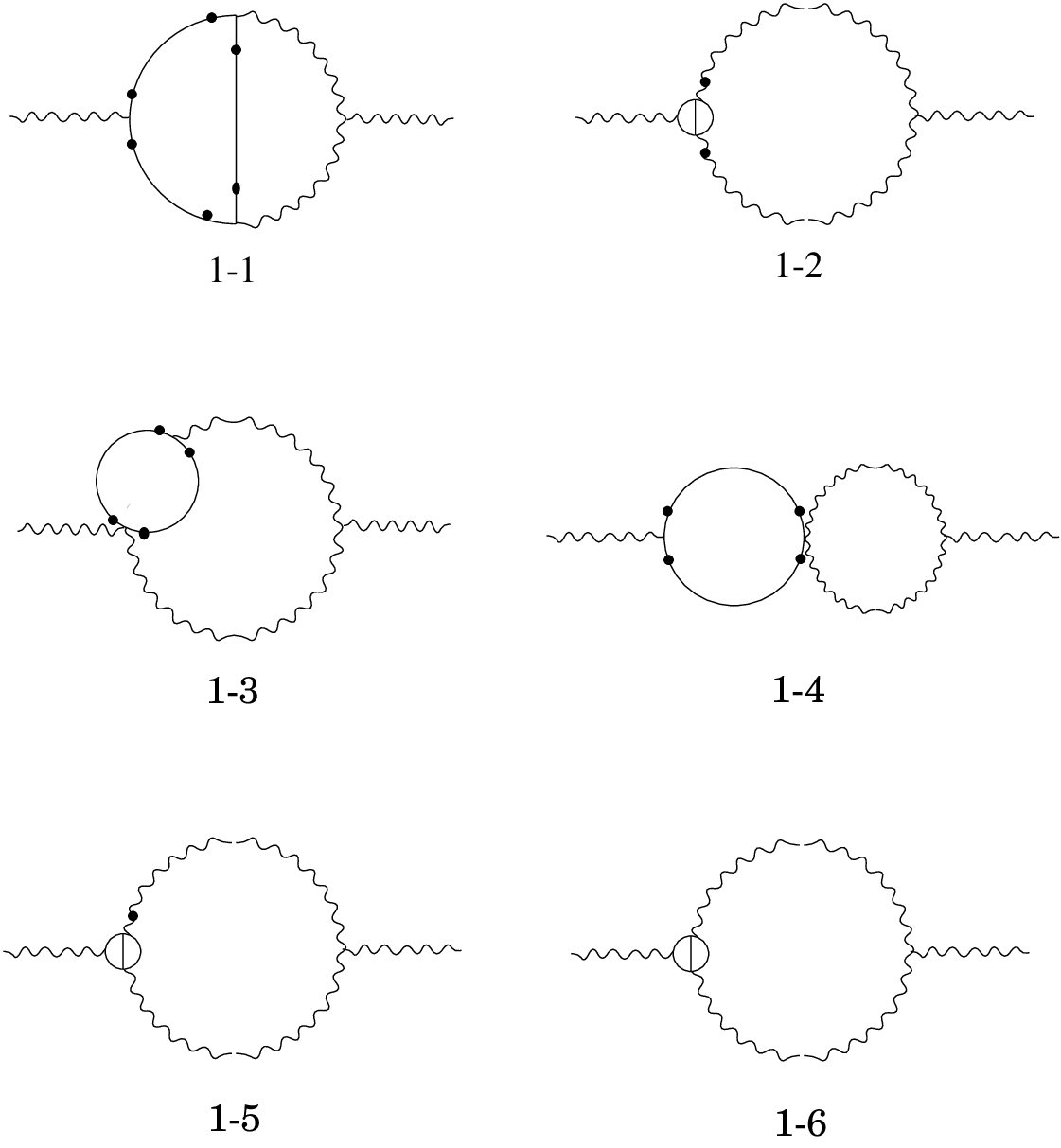 } }
 \caption{ Diagrams \ of \ Group \ 1}
 \label{}
\end{figure}

\begin{table}
\begin{center}
 \doublerulesep=0pt
 \def\arraystretch{1.7}
 \begin{tabular}{|c|c|c|c|}  \hline\hline
 \multicolumn{1}{|c|}{Diagram} & \multicolumn{1}{c|}{$A$} &
                                      \multicolumn{1}{c|}{$B$} &
                                 \multicolumn{1}{c|}{$C$} \\ \hline\hline
  $ 1-1$ & $ -\frac{1}{4\epsilon^2}+(\frac{13}{48}-\frac{\gamma}{4}
           -\frac{\rho}{4}-\sigma)\frac{1}{\epsilon}$&
           $ \frac{1}{8\epsilon^2}+(-\frac{13}{96}+\frac{\gamma}{8}
           +\frac{\rho}{8}+\frac{\sigma}{2})\frac{1}{\epsilon}$&
           $0$\\ \hline
  $ 1-2$ & $ \frac{1}{2\epsilon^2}+(-\frac{4}{3}+\frac{\gamma}{4}
           +\frac{\rho}{4})\frac{1}{\epsilon}$&
           $ -\frac{1}{4\epsilon^2}+(\frac{2}{3}-\frac{\gamma}{8}
           -\frac{\rho}{8})\frac{1}{\epsilon}$&
           $0$\\ \hline
  $ 1-3$ & $ (\frac{13}{12}
           +\frac{\sigma}{3})\frac{1}{\epsilon}$&
           $(-\frac{13}{24}
           -\frac{\sigma}{6})\frac{1}{\epsilon}$&
           $0$\\ \hline
  $ 1-4$ & $ \frac{\sigma}{2\epsilon}$&
           $-\frac{\sigma}{4\epsilon}$&
           $0$\\ \hline
  $ 1-5$ & $-\frac{\sigma}{2\epsilon}$&
           $\frac{\sigma}{4\epsilon}$&
           $\frac{\sigma}{3\epsilon}$\\ \hline
  $ 1-6$ & $ \frac{2\sigma}{3\epsilon}$&
           $-\frac{\sigma}{3\epsilon}$&
           $-\frac{\sigma}{3\epsilon}$\\ \hline\hline
 \multicolumn{1}{|c|}{Total} & \multicolumn{1}{c|}
                   { $\frac{1}{4\epsilon^2}+\frac{1}{48\epsilon}$ }&
                   \multicolumn{1}{c|}
                   { $-\frac{1}{8\epsilon^2}-\frac{1}{96\epsilon}$ } &
                   \multicolumn{1}{c|}{$0$}\\ \hline\hline
 \end{tabular}
\end{center}
\caption{ The results for Group $1$}
\label{Group1}
\end{table}%

\begin{figure}
 \leavevmode
 \epsfxsize = 10 cm
 \centerline{ \epsfbox{ 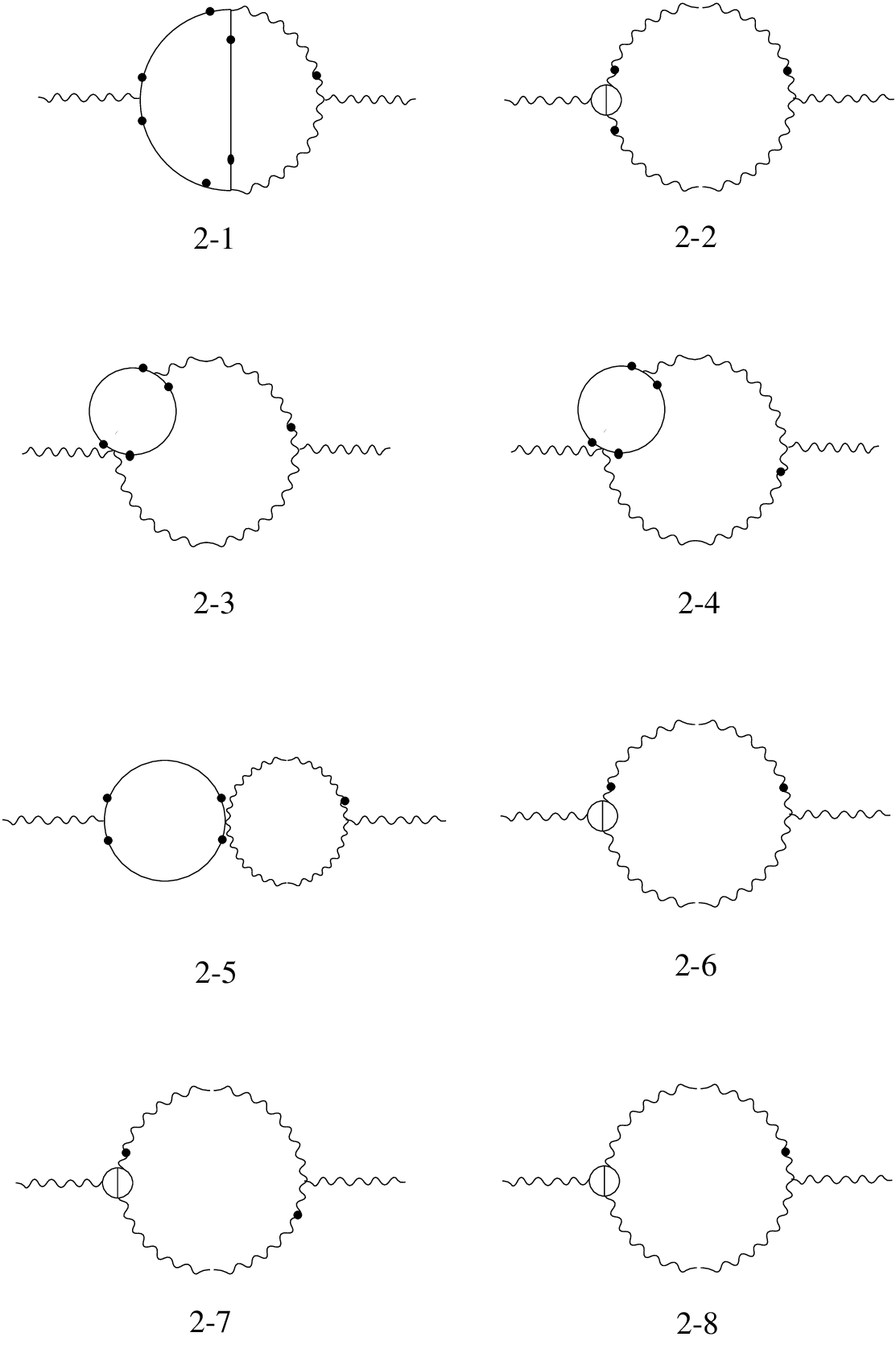 } }
 \caption{ Diagrams \ of \ Group \ 2}
 \label{}
\end{figure}

\begin{table}
\begin{center}
 \doublerulesep=0pt
 \def\arraystretch{1.7}
 \begin{tabular}{|c|c|c|c|}  \hline\hline
  \multicolumn{1}{|c|}{Diagram} & \multicolumn{1}{c|}{$A$} &
                                        \multicolumn{1}{c|}{$B$} &
                                 \multicolumn{1}{c|}{$C$} \\ \hline\hline
  $ 2-1$ & $ -\frac{1}{4\epsilon^2}+(\frac{5}{16}-\frac{\gamma}{4}
           -\frac{\rho}{4})\frac{1}{\epsilon}$&
           $ \frac{1}{8\epsilon^2}+(-\frac{7}{32}+\frac{\gamma}{8}
           +\frac{\rho}{8})\frac{1}{\epsilon}$&
           $0$\\ \hline
  $ 2-2$ & $ -\frac{5}{6\epsilon^2}+(\frac{5}{12}-\frac{5\gamma}{12}
           -\frac{5\rho}{12})\frac{1}{\epsilon}$&
           $ \frac{5}{12\epsilon^2}+(-\frac{1}{12}+\frac{5\gamma}{24}
           +\frac{5\rho}{24})\frac{1}{\epsilon}$&
           $0$\\ \hline
  $ 2-3$ & $ \frac{1}{4\epsilon^2}+(-\frac{7}{48}+\frac{\gamma}{4}
           +\frac{\rho}{4}+\frac{\sigma}{2})\frac{1}{\epsilon}$&
           $ -\frac{1}{8\epsilon^2}+(\frac{13}{96}-\frac{\gamma}{8}
           -\frac{\rho}{8}-\frac{\sigma}{4})\frac{1}{\epsilon}$&
           $0$\\ \hline
  $ 2-4$ & $ -\frac{1}{4\epsilon^2}+(\frac{31}{48}-\frac{\gamma}{4}
           -\frac{\rho}{4})\frac{1}{\epsilon}$&
           $ \frac{1}{8\epsilon^2}+(-\frac{37}{96}+\frac{\gamma}{8}
           +\frac{\rho}{8})\frac{1}{\epsilon}$&
           $0$\\ \hline
  $ 2-5$ & $0$&
           $0$&
           $0$\\ \hline
  $ 2-6$ & $ \frac{2}{3\epsilon^2}+(-\frac{1}{6}+\frac{\gamma}{3}
           +\frac{\rho}{3})\frac{1}{\epsilon}$&
           $ -\frac{1}{3\epsilon^2}+(\frac{1}{8}-\frac{\gamma}{6}
           -\frac{\rho}{6})\frac{1}{\epsilon}$&
           $0$\\ \hline
  $ 2-7$ & $ \frac{2}{3\epsilon^2}+(-\frac{5}{6}+\frac{\gamma}{3}
           +\frac{\rho}{3})\frac{1}{\epsilon}$&
           $ -\frac{1}{3\epsilon^2}+(\frac{11}{24}-\frac{\gamma}{6}
           -\frac{\rho}{6})\frac{1}{\epsilon}$&
           $0$\\ \hline
  $ 2-8$ & $-\frac{\sigma}{2\epsilon}$&
           $\frac{\sigma}{4\epsilon}$&
           $0$\\ \hline\hline
 \multicolumn{1}{|c|}{Total} & \multicolumn{1}{c|}
                   { $\frac{1}{4\epsilon^2}+\frac{11}{48\epsilon}$ }&
                   \multicolumn{1}{c|}
                   { $-\frac{1}{8\epsilon^2}+\frac{1}{32\epsilon}$ } &
                   \multicolumn{1}{c|}{$0$}\\ \hline\hline
 \end{tabular}
\end{center}
\caption{ The results for Group $2$}
\label{Group2}
\end{table}%

\begin{figure}
 \leavevmode
 \epsfxsize = 10 cm
 \centerline{ \epsfbox{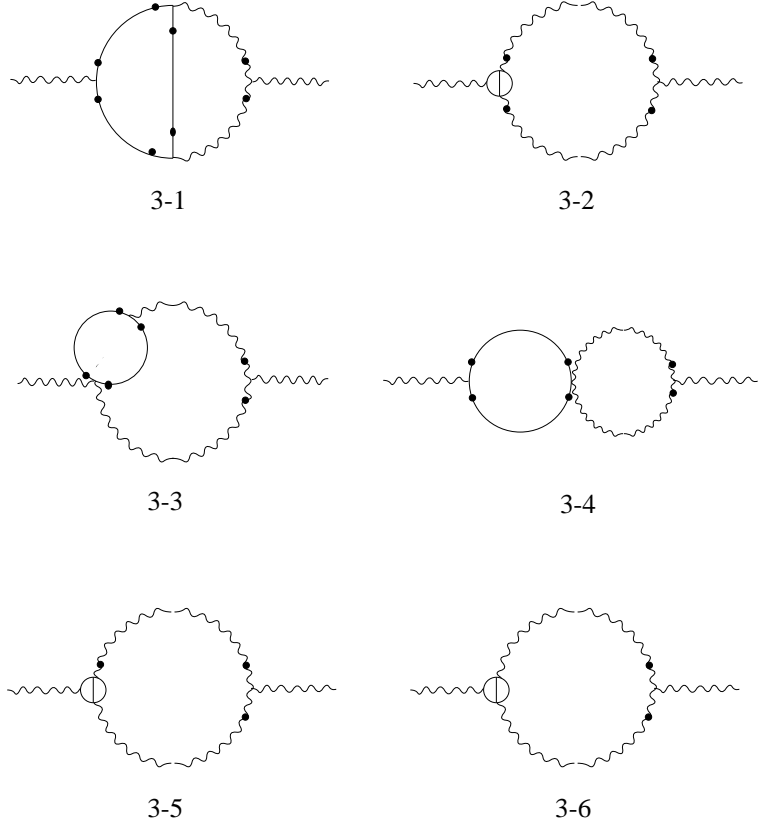} }
 \caption{ Diagrams \ of \ Group \ 3}
 \label{}
\end{figure}

\begin{table}
\begin{center}
 \doublerulesep=0pt
 \def\arraystretch{1.7}
 \begin{tabular}{|c|c|c|c|}  \hline\hline
  \multicolumn{1}{|c|}{Diagram} & \multicolumn{1}{c|}{$A$} &
                                        \multicolumn{1}{c|}{$B$} &
                                 \multicolumn{1}{c|}{$C$} \\ \hline\hline
  $ 3-1$ & $\frac{19}{24\epsilon}$&
           $-\frac{11}{24\epsilon}$&
           $0$\\ \hline
  $ 3-2$ & $\frac{11}{36\epsilon}$&
           $-\frac{5}{18\epsilon}$&
           $0$\\ \hline
  $ 3-3$ & $ \frac{1}{2\epsilon^2}+(-\frac{4}{3}+\frac{\gamma}{2}
           +\frac{\rho}{2})\frac{1}{\epsilon}$&
           $-\frac{1}{4\epsilon^2}+(\frac{19}{24}-\frac{\gamma}{4}
           -\frac{\rho}{4})\frac{1}{\epsilon}$&
           $0$\\ \hline
  $ 3-4$ & $-\frac{1}{2\epsilon}$&
           $\frac{1}{4\epsilon}$&
           $0$\\ \hline
  $ 3-5$ & $\frac{2}{3\epsilon}$&
           $-\frac{1}{3\epsilon}$& $\frac{1}{6\epsilon}$\\ \hline
  $ 3-6$ & $ -\frac{1}{\epsilon^2}+(\frac{1}{3}-\frac{\gamma}{2}
           -\frac{\rho}{2})\frac{1}{\epsilon}$&
           $\frac{1}{2\epsilon^2}+(-\frac{1}{6}+\frac{\gamma}{4}
           +\frac{\rho}{4})\frac{1}{\epsilon}$&
           $-\frac{1}{6\epsilon}$\\ \hline\hline
 \multicolumn{1}{|c|}{Total} & \multicolumn{1}{c|}
                   { $-\frac{1}{2\epsilon^2}+\frac{19}{72\epsilon}$ }&
                   \multicolumn{1}{c|}
                   { $\frac{1}{4\epsilon^2}-\frac{7}{36\epsilon}$ } &
                   \multicolumn{1}{c|}{$0$}\\ \hline\hline
 \end{tabular}
\end{center}
\caption{ The results for Group $3$}
\label{Group3}
\end{table}%

\begin{figure}
 \leavevmode
 \epsfxsize = 10 cm
 \centerline{ \epsfbox{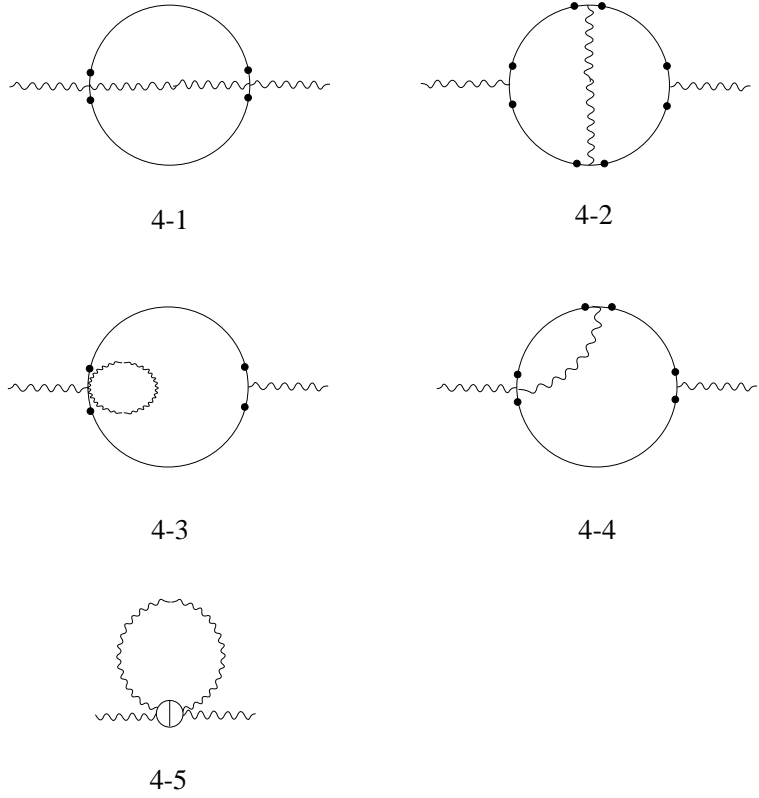} }
 \caption{ Diagrams \ of \ Group \ 4}
 \label{}
\end{figure}

\begin{table}
\begin{center}
 \doublerulesep=0pt
 \def\arraystretch{1.7}
 \begin{tabular}{|c|c|c|c|}  \hline\hline
  \multicolumn{1}{|c|}{Diagram} & \multicolumn{1}{c|}{$A$} &
                                        \multicolumn{1}{c|}{$B$} &
                                 \multicolumn{1}{c|}{$C$} \\ \hline\hline
  $ 4-1$ & $ -\frac{1}{2\epsilon^2}+(\frac{13}{24}-\frac{\gamma}{2}
           -\frac{\rho}{2}-\frac{\sigma}{2})\frac{1}{\epsilon}$&
           $ \frac{1}{8\epsilon^2}+(-\frac{19}{48}+\frac{\gamma}{8}
           +\frac{\rho}{8}+\frac{\sigma}{8})\frac{1}{\epsilon}$&
           $0$\\ \hline
  $ 4-2$ & $ \frac{1}{4\epsilon^2}+(-\frac{1}{16}+\frac{\gamma}{4}
           +\frac{\rho}{4}+\frac{\sigma}{4})\frac{1}{\epsilon}$&
           $ -\frac{1}{4\epsilon^2}+(\frac{1}{32}-\frac{\gamma}{4}
           -\frac{\rho}{4}-\frac{\sigma}{4})\frac{1}{\epsilon}$&
           $0$\\ \hline
  $ 4-3$ & $ \frac{1}{3\epsilon^2}+(\frac{11}{36}+\frac{\gamma}{3}
           +\frac{\rho}{3}+\frac{\sigma}{6})\frac{1}{\epsilon}$&
           $ -\frac{1}{6\epsilon^2}+(-\frac{11}{72}-\frac{\gamma}{6}
           -\frac{\rho}{6}-\frac{\sigma}{12})\frac{1}{\epsilon}$&
           $-\frac{1}{6\epsilon}$\\ \hline
  $ 4-4$ & $ -\frac{1}{3\epsilon^2}+(-\frac{13}{72}-\frac{\gamma}{3}
           -\frac{\rho}{3}-\frac{\sigma}{6})\frac{1}{\epsilon}$&
           $ \frac{5}{12\epsilon^2}+(\frac{31}{144}+\frac{5\gamma}{12}
           +\frac{5\rho}{12}+\frac{\sigma}{3})\frac{1}{\epsilon}$&
           $\frac{1}{6\epsilon}$\\ \hline
  $ 4-5$ & $ \frac{1}{2\epsilon^2}+(-\frac{7}{8}+\frac{\gamma}{4}
           +\frac{\rho}{4}+\frac{\sigma}{4})\frac{1}{\epsilon}$&
           $ -\frac{1}{4\epsilon^2}+(\frac{19}{48}-\frac{\gamma}{8}
           -\frac{\rho}{8}-\frac{\sigma}{8})\frac{1}{\epsilon}$&
           $0$\\ \hline\hline
 \multicolumn{1}{|c|}{Total} & \multicolumn{1}{c|}
                   { $\frac{1}{4\epsilon^2}-\frac{13}{48\epsilon}$ }&
                   \multicolumn{1}{c|}
                   { $-\frac{1}{8\epsilon^2}+\frac{3}{32\epsilon}$ } &
                   \multicolumn{1}{c|}{$0$}\\ \hline\hline
 \end{tabular}
\end{center}
\caption{ The results for Group $4$}
\label{Group4}
\end{table}%

\begin{figure}
 \leavevmode
 \epsfxsize = 10 cm
 \centerline{ \epsfbox{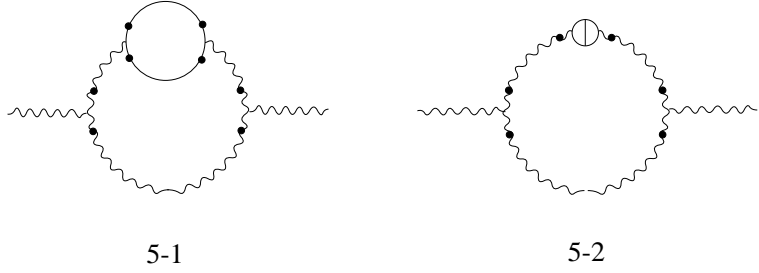} }
 \caption{ Diagrams \ of \ Group \ 5}
 \label{}
\end{figure}

\begin{table}
\begin{center}
 \doublerulesep=0pt
 \def\arraystretch{1.7}
 \begin{tabular}{|c|c|c|c|}  \hline\hline
  \multicolumn{1}{|c|}{Diagram} & \multicolumn{1}{c|}{$A$} &
                                        \multicolumn{1}{c|}{$B$} &
                                 \multicolumn{1}{c|}{$C$} \\ \hline\hline
  $ 5-1$ & $ -\frac{1}{8\epsilon^2}+(\frac{13}{32}-\frac{\gamma}{8}
           -\frac{\rho}{8})\frac{1}{\epsilon}$&
           $ \frac{1}{16\epsilon^2}+(-\frac{19}{64}+\frac{\gamma}{16}
           +\frac{\rho}{16})\frac{1}{\epsilon}$&
           $0$\\ \hline
  $ 5-2$ & $ \frac{1}{4\epsilon^2}+(-\frac{17}{36}+\frac{\gamma}{8}
           +\frac{\rho}{8})\frac{1}{\epsilon}$&
           $ -\frac{1}{8\epsilon^2}+(\frac{61}{144}-\frac{\gamma}{16}
           -\frac{\rho}{16})\frac{1}{\epsilon}$&
           $0$\\ \hline\hline
 \multicolumn{1}{|c|}{Total} & \multicolumn{1}{c|}
                   { $\frac{1}{8\epsilon^2}-\frac{19}{288\epsilon}$ }&
                   \multicolumn{1}{c|}
                   { $-\frac{1}{16\epsilon^2}+\frac{73}{576\epsilon}$ } &
                   \multicolumn{1}{c|}{$0$}\\ \hline\hline
 \end{tabular}
\end{center}
\caption{ The results for Group $5$}
\label{Group5}
\end{table}%

\begin{figure}
 \leavevmode
 \epsfxsize = 10 cm
 \centerline{ \epsfbox{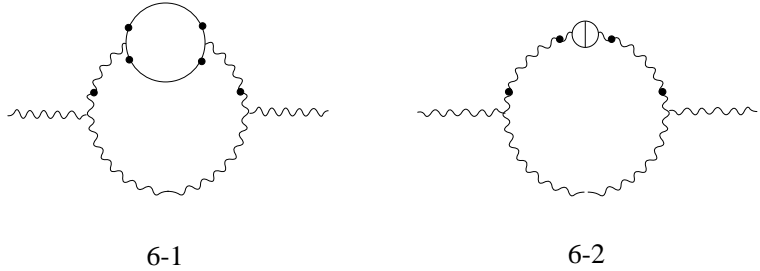} }
 \caption{ Diagrams \ of \ Group \ 6}
 \label{}
\end{figure}

\begin{table}
\begin{center}
 \doublerulesep=0pt
 \def\arraystretch{1.7}
 \begin{tabular}{|c|c|c|c|}  \hline\hline
  \multicolumn{1}{|c|}{Diagram} & \multicolumn{1}{c|}{$A$} &
                                        \multicolumn{1}{c|}{$B$} &
                                 \multicolumn{1}{c|}{$C$} \\ \hline\hline
  $ 6-1$ & $ (-\frac{13}{192}-\frac{\sigma}{8})\frac{1}{\epsilon}$&
           $ (\frac{7}{384}+\frac{\sigma}{16})\frac{1}{\epsilon}$&
           $0$\\ \hline
  $ 6-2$ & $ (\frac{1}{48}+\frac{\sigma}{8})\frac{1}{\epsilon}$&
           $ -\frac{\sigma}{16}$&
           $0$\\ \hline\hline
 \multicolumn{1}{|c|}{Total} & \multicolumn{1}{c|}
                   { $-\frac{3}{64\epsilon}$ }&
                   \multicolumn{1}{c|}
                   { $\frac{7}{384\epsilon}$ } &
                   \multicolumn{1}{c|}{$0$}\\ \hline\hline
 \end{tabular}
\end{center}
\caption{ The results for Group $6$}
\label{Group6}
\end{table}%

\begin{figure}
 \leavevmode
 \epsfxsize = 10 cm
 \centerline{ \epsfbox{ 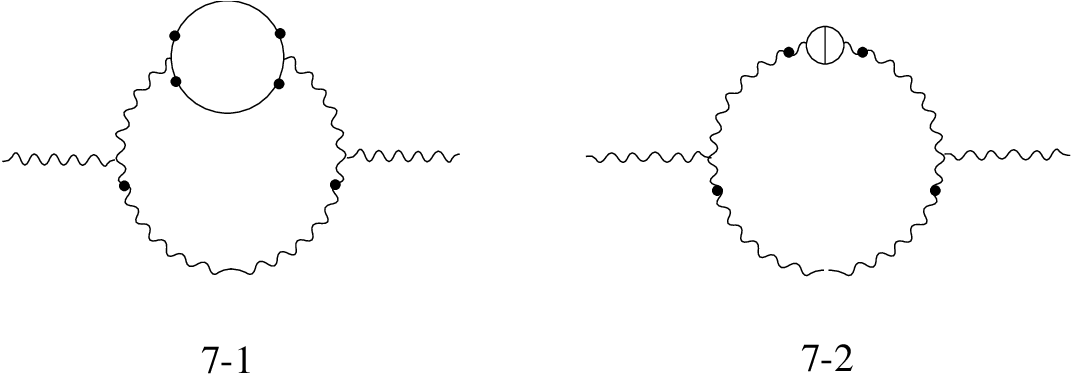 } }
 \caption{ Diagrams \ of \ Group \ 7}
 \label{}
\end{figure}

\begin{table}
\begin{center}
 \doublerulesep=0pt
 \def\arraystretch{1.7}
 \begin{tabular}{|c|c|c|c|}  \hline\hline
  \multicolumn{1}{|c|}{Diagram} & \multicolumn{1}{c|}{$A$} &
                                        \multicolumn{1}{c|}{$B$} &
                                 \multicolumn{1}{c|}{$C$} \\ \hline\hline
  $ 7-1$ & $\frac{11}{192\epsilon}$&
           $-\frac{29}{384\epsilon}$&
           $0$\\ \hline
  $ 7-2$ & $-\frac{5}{48\epsilon}$&
           $ \frac{1}{8\epsilon}$&
           $0$\\ \hline\hline
 \multicolumn{1}{|c|}{Total} & \multicolumn{1}{c|}
                   { $-\frac{3}{64\epsilon}$ }&
                   \multicolumn{1}{c|}
                   { $\frac{19}{384\epsilon}$ } &
                   \multicolumn{1}{c|}{$0$}\\ \hline\hline
 \end{tabular}
\end{center}
\caption{ The results for Group $7$}
\label{Group7}
\end{table}%

\begin{figure}
 \leavevmode
 \epsfxsize = 10 cm
 \centerline{ \epsfbox{ 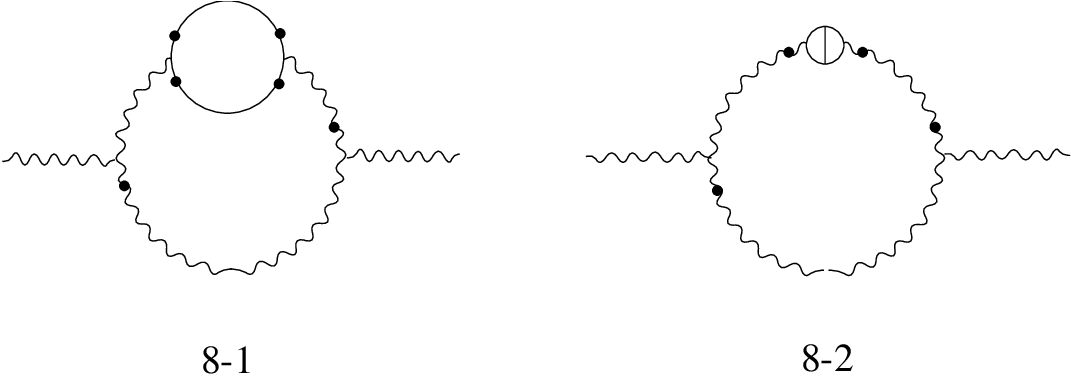 } }
 \caption{ Diagrams \ of \ Group \ 8}
 \label{}
\end{figure}

\begin{table}
\begin{center}
 \doublerulesep=0pt
 \def\arraystretch{1.7}
 \begin{tabular}{|c|c|c|c|}  \hline\hline
  \multicolumn{1}{|c|}{Diagram} & \multicolumn{1}{c|}{$A$} &
                                        \multicolumn{1}{c|}{$B$} &
                                 \multicolumn{1}{c|}{$C$} \\ \hline\hline
  $ 8-1$ & $ \frac{1}{8\epsilon^2}+(-\frac{13}{48}+\frac{\gamma}{8}
           +\frac{\rho}{8})\frac{1}{\epsilon}$&
           $ -\frac{1}{16\epsilon^2}+(\frac{5}{48}-\frac{\gamma}{16}
           -\frac{\rho}{16})\frac{1}{\epsilon}$&
           $0$\\ \hline
  $ 8-2$ & $ -\frac{1}{4\epsilon^2}+(\frac{5}{24}-\frac{\gamma}{8}
           -\frac{\rho}{8})\frac{1}{\epsilon}$&
           $ \frac{1}{8\epsilon^2}+(-\frac{1}{12}+\frac{\gamma}{16}
           +\frac{\rho}{16})\frac{1}{\epsilon}$&
           $0$\\ \hline\hline
 \multicolumn{1}{|c|}{Total} & \multicolumn{1}{c|}
                   { $-\frac{1}{8\epsilon^2}-\frac{1}{16\epsilon}$ }&
                   \multicolumn{1}{c|}
                   { $\frac{1}{16\epsilon^2}+\frac{1}{48\epsilon}$ } &
                   \multicolumn{1}{c|}{$0$}\\ \hline\hline
 \end{tabular}
\end{center}
\caption{ The results for Group $8$}
\label{Group8}
\end{table}%

$ \ $
\newpage

\begin{figure}
 \leavevmode
 \epsfxsize = 10 cm
 \centerline{ \epsfbox{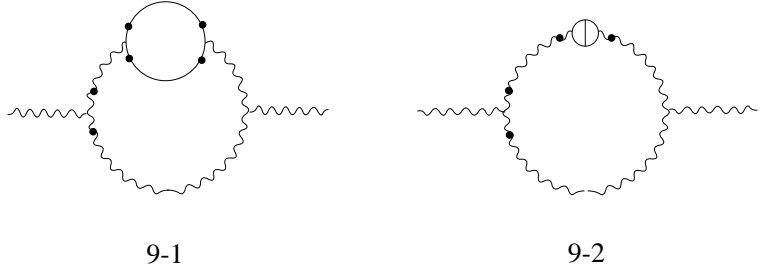} }
 \caption{ Diagrams \ of \ Group \ 9}
 \label{}
\end{figure}

\begin{table}
\begin{center}
 \doublerulesep=0pt
 \def\arraystretch{1.7}
 \begin{tabular}{|c|c|c|c|}  \hline\hline
  \multicolumn{1}{|c|}{Diagram} & \multicolumn{1}{c|}{$A$} &
                                        \multicolumn{1}{c|}{$B$} &
                                 \multicolumn{1}{c|}{$C$} \\ \hline\hline
  $ 9-1$ & $ \frac{1}{4\epsilon^2}+(-\frac{25}{16}+\frac{\gamma}{4}
           +\frac{\rho}{4})\frac{1}{\epsilon}$&
           $ -\frac{1}{8\epsilon^2}+(\frac{25}{32}-\frac{\gamma}{8}
           -\frac{\rho}{8})\frac{1}{\epsilon}$&
           $0$\\ \hline
  $ 9-2$ & $ -\frac{1}{2\epsilon^2}+(\frac{5}{3}-\frac{\gamma}{4}
           -\frac{\rho}{4})\frac{1}{\epsilon}$&
           $ \frac{1}{4\epsilon^2}+(-\frac{5}{6}+\frac{\gamma}{8}
           +\frac{\rho}{8})\frac{1}{\epsilon}$&
           $0$\\ \hline\hline
 \multicolumn{1}{|c|}{Total} & \multicolumn{1}{c|}
                   { $-\frac{1}{4\epsilon^2}+\frac{5}{48\epsilon}$ }&
                   \multicolumn{1}{c|}
                   { $\frac{1}{8\epsilon^2}-\frac{5}{96\epsilon}$ } &
                   \multicolumn{1}{c|}{$0$}\\ \hline\hline
 \end{tabular}
\end{center}
\caption{ The results for Group $9$}
\label{Group9}
\end{table}%

\begin{figure}
 \leavevmode
 \epsfxsize = 10 cm
 \centerline{ \epsfbox{ 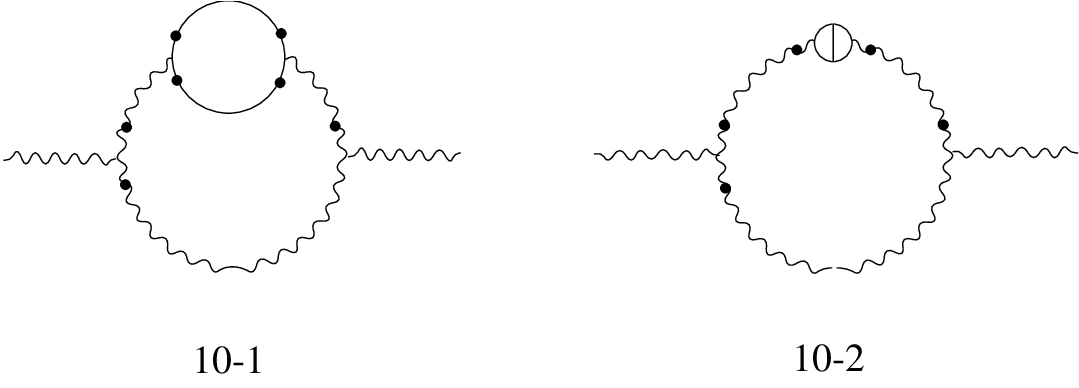 } }
 \caption{ Diagrams \ of \ Group \ 10}
 \label{}
\end{figure}

\begin{table}
\begin{center}
 \doublerulesep=0pt
 \def\arraystretch{1.7}
 \begin{tabular}{|c|c|c|c|}  \hline\hline
  \multicolumn{1}{|c|}{Diagram} & \multicolumn{1}{c|}{$A$} &
                                        \multicolumn{1}{c|}{$B$} &
                                 \multicolumn{1}{c|}{$C$} \\ \hline\hline
  $ 10-1$ & $ -\frac{1}{8\epsilon^2}+(\frac{13}{32}-\frac{\gamma}{8}
           -\frac{\rho}{8})\frac{1}{\epsilon}$&
           $ \frac{1}{16\epsilon^2}+(-\frac{11}{64}+\frac{\gamma}{16}
           +\frac{\rho}{16})\frac{1}{\epsilon}$&
           $0$\\ \hline
  $ 10-2$ & $ \frac{1}{4\epsilon^2}+(-\frac{11}{24}+\frac{\gamma}{8}
           +\frac{\rho}{8})\frac{1}{\epsilon}$&
           $ -\frac{1}{8\epsilon^2}+(\frac{1}{6}-\frac{\gamma}{16}
           -\frac{\rho}{16})\frac{1}{\epsilon}$&
           $0$\\ \hline\hline
 \multicolumn{1}{|c|}{Total} & \multicolumn{1}{c|}
                   { $\frac{1}{8\epsilon^2}-\frac{5}{96\epsilon}$ }&
                   \multicolumn{1}{c|}
                   { $-\frac{1}{16\epsilon^2}-\frac{1}{192\epsilon}$ } &
                   \multicolumn{1}{c|}{$0$}\\ \hline\hline
 \end{tabular}
\end{center}
\caption{ The results for Group $10$}
\label{Group10}
\end{table}%

\begin{figure}
 \leavevmode
 \epsfxsize = 10 cm
 \centerline{ \epsfbox{ 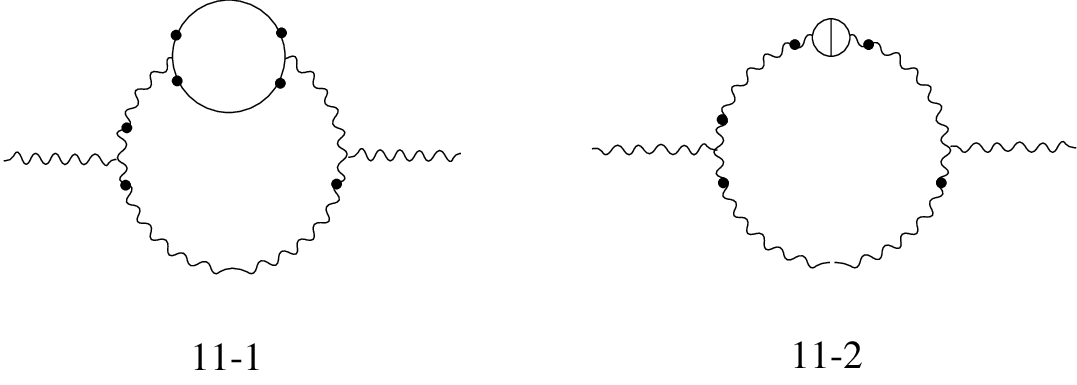 } }
 \caption{ Diagrams \ of \ Group \ 11}
 \label{}
\end{figure}

\begin{table}
\begin{center}
 \doublerulesep=0pt
 \def\arraystretch{1.7}
 \begin{tabular}{|c|c|c|c|}  \hline\hline
  \multicolumn{1}{|c|}{Diagram} & \multicolumn{1}{c|}{$A$} &
                                        \multicolumn{1}{c|}{$B$} &
                                 \multicolumn{1}{c|}{$C$} \\ \hline\hline
  $ 11-1$ & $ \frac{1}{8\epsilon^2}+(-\frac{13}{32}+\frac{\gamma}{8}
           +\frac{\rho}{8})\frac{1}{\epsilon}$&
           $ -\frac{1}{16\epsilon^2}+(\frac{19}{64}-\frac{\gamma}{16}
           -\frac{\rho}{16})\frac{1}{\epsilon}$&
           $0$\\ \hline
  $ 11-2$ & $ -\frac{1}{4\epsilon^2}+(\frac{11}{24}-\frac{\gamma}{8}
           -\frac{\rho}{8})\frac{1}{\epsilon}$&
           $ \frac{1}{8\epsilon^2}+(-\frac{5}{12}+\frac{\gamma}{16}
           +\frac{\rho}{16})\frac{1}{\epsilon}$&
           $0$\\ \hline\hline
 \multicolumn{1}{|c|}{Total} & \multicolumn{1}{c|}
                   { $-\frac{1}{8\epsilon^2}+\frac{5}{96\epsilon}$ }&
                   \multicolumn{1}{c|}
                   { $\frac{1}{16\epsilon^2}-\frac{23}{192\epsilon}$ } &
                   \multicolumn{1}{c|}{$0$}\\ \hline\hline
 \end{tabular}
\end{center}
\caption{ The results for Group $11$}
\label{Group11}
\end{table}%

\begin{figure}
 \leavevmode
 \epsfxsize = 10 cm
 \centerline{ \epsfbox{ 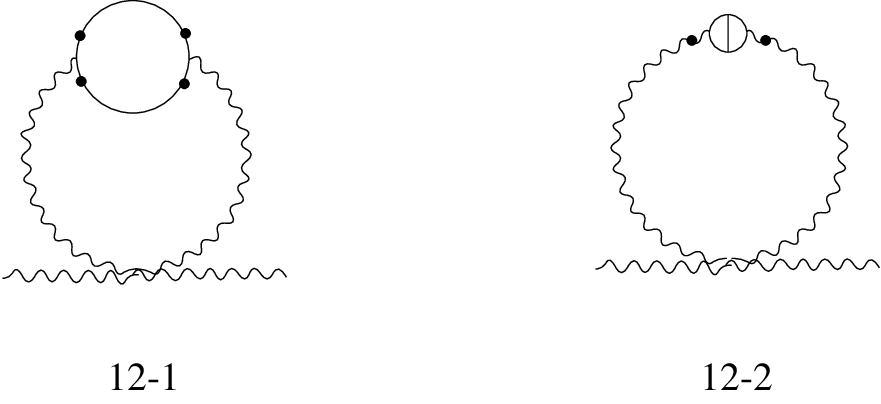 } }
 \caption{ Diagrams \ of \ Group \ 12}
 \label{}
\end{figure}

\begin{table}
\begin{center}
 \doublerulesep=0pt
 \def\arraystretch{1.7}
 \begin{tabular}{|c|c|c|c|}  \hline\hline
  \multicolumn{1}{|c|}{Diagram} & \multicolumn{1}{c|}{$A$} &
                                        \multicolumn{1}{c|}{$B$} &
                                 \multicolumn{1}{c|}{$C$} \\ \hline\hline
  $ 12-1$ & $ \frac{5}{16\epsilon}$&
            $ -\frac{3}{32\epsilon}$&
           $0$\\ \hline
  $ 12-2$ & $ -\frac{5}{12\epsilon}$&
           $ \frac{1}{8\epsilon}$&
           $0$\\ \hline\hline
 \multicolumn{1}{|c|}{Total} & \multicolumn{1}{c|}
                   { $-\frac{5}{48\epsilon}$ }&
                   \multicolumn{1}{c|}
                   { $\frac{1}{32\epsilon}$} &
                   \multicolumn{1}{c|}{$0$}\\ \hline\hline
 \end{tabular}
\end{center}
\caption{ The results for Group $12$}
\label{Group 12}
\end{table}%

\begin{figure}
 \leavevmode
 \epsfxsize = 10 cm
 \centerline{ \epsfbox{ 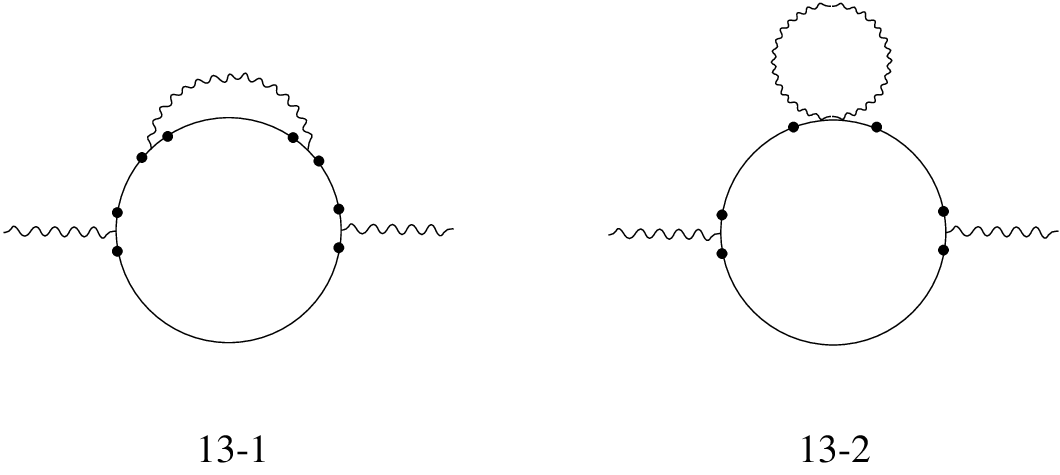 } }
 \caption{ Diagrams \ of \ Group \ 13}
 \label{}
\end{figure}

\begin{table}
\begin{center}
 \doublerulesep=0pt
 \def\arraystretch{1.7}
 \begin{tabular}{|c|c|c|c|}  \hline\hline
  \multicolumn{1}{|c|}{Diagram} & \multicolumn{1}{c|}{$A$} &
                                        \multicolumn{1}{c|}{$B$} &
                                 \multicolumn{1}{c|}{$C$} \\ \hline\hline
  $ 13-1$ & $ \frac{1}{3\epsilon^2}+(-\frac{23}{72}+\frac{\gamma}{3}
           +\frac{\rho}{3}+\frac{\sigma}{6})\frac{1}{\epsilon}$&
           $ -\frac{1}{6\epsilon^2}+(\frac{23}{144}-\frac{\gamma}{6}
           -\frac{\rho}{6}-\frac{\sigma}{12})\frac{1}{\epsilon}$&
           $-\frac{1}{6\epsilon}$\\ \hline
  $ 13-2$ & $ -\frac{1}{3\epsilon^2}+(\frac{7}{36}-\frac{\gamma}{3}
           -\frac{\rho}{3}-\frac{\sigma}{6})\frac{1}{\epsilon}$&
           $ \frac{1}{6\epsilon^2}+(-\frac{7}{72}+\frac{\gamma}{6}
           +\frac{\rho}{6}+\frac{\sigma}{12})\frac{1}{\epsilon}$&
           $\frac{1}{6\epsilon}$\\ \hline\hline
 \multicolumn{1}{|c|}{Total} & \multicolumn{1}{c|}
                   { $-\frac{1}{8\epsilon}$ }&
                   \multicolumn{1}{c|}
                   { $\frac{1}{16\epsilon}$ } &
                   \multicolumn{1}{c|}{$0$}\\ \hline\hline
 \end{tabular}
\end{center}
\caption{ The results for Group $13$}
\label{Group 13}
\end{table}%

\begin{table}
\begin{center}
 \doublerulesep=0pt
 \def\arraystretch{1.7}
 \begin{tabular}{|c|c|c|}  \hline\hline
  \multicolumn{1}{|c|}{Group} & \multicolumn{1}{c|}{$A$} &
                                        \multicolumn{1}{c|}{$B$} \\
\hline\hline
  $ 1$ & $ {1 \over 4\epsilon^2}+{ 1 \over 48\epsilon} $ &
                     $ -{1 \over 8\epsilon^2}-{1 \over  96\epsilon} $ \\ \hline
  $ 2$ & $ {1 \over 4\epsilon^2}+{11 \over 48\epsilon} $ &
                     $ -{1 \over 8\epsilon^2}+{1 \over  32\epsilon} $ \\ \hline
  $ 3$ & $-{1 \over 2\epsilon^2}+{19 \over 72\epsilon} $ &
                     $  {1 \over 4\epsilon^2}-{7 \over  36\epsilon} $ \\ \hline
  $ 4$ & $ {1 \over 4\epsilon^2}-{13 \over 48\epsilon} $ &
                     $ -{1 \over 8\epsilon^2}+{3 \over  32\epsilon} $ \\ \hline
  $ 5$ & $ {1 \over 8\epsilon^2}-{19 \over 288\epsilon} $ &
                  $ -{1 \over 16\epsilon^2}+{73 \over  576\epsilon} $ \\ \hline
  $ 6$ & $ -{3 \over 64\epsilon} $ &
                  $ {7 \over  384\epsilon} $ \\ \hline
  $ 7$ & $ -{3 \over 64\epsilon} $ &
                 $ {19 \over  384\epsilon} $ \\ \hline
  $ 8$ & $ -{1 \over 8\epsilon^2}-{1 \over 16\epsilon} $ &
                   $ {1 \over 16\epsilon^2}+{1 \over 48\epsilon} $ \\ \hline
 $ 9$ & $-{1 \over 4\epsilon^2}+{ 5 \over 48\epsilon} $ &
                  $  {1 \over 8\epsilon^2}-{5 \over  96\epsilon} $ \\ \hline
  $ 10$ & $ {1 \over 8\epsilon^2}-{5 \over 96\epsilon} $ &
               $ -{1 \over 16\epsilon^2}-{1 \over  192\epsilon} $ \\ \hline
  $ 11$ & $ -{1 \over 8\epsilon^2}+{5 \over 96\epsilon} $ &
                $ {1 \over 16\epsilon^2}-{23 \over  192\epsilon} $ \\ \hline
  $12$ & $ -{ 5 \over 48\epsilon} $ & $ {1 \over  32\epsilon} $ \\ \hline
  $13$ & $ -{ 1 \over  8\epsilon} $ & $ {1 \over  16\epsilon} $ \\ \hline\hline
  \multicolumn{1}{|c|}{Total} & \multicolumn{1}{c|}{ $-{5 \over 48\epsilon}$ }&
                   \multicolumn{1}{c|}{ ${5 \over 96\epsilon}$ } \\
\hline\hline
 \end{tabular}
\end{center}
\caption{Divergences for $\hat{R}$}
\label{result}
\end{table}
%

%

%
%
%
%
%
%



\end{document}